\documentclass[%
 reprint,
 amsmath,amssymb,
 aps,
floatfix,
]{revtex4-2}

\usepackage{amsmath}
\usepackage{amsthm}
\usepackage{amssymb}
\usepackage{mathtools}
\usepackage{hyperref}
\usepackage{cleveref}
\usepackage{ifthen}
\usepackage{xcolor}
\usepackage{lipsum}  
\usepackage{xspace}
\usepackage{multirow}
\usepackage{graphicx,ragged2e}
\usepackage[caption=false]{subfig}
\usepackage{xr}

\definecolor{dartmouthgreen}{RGB}{0, 105, 62}

\newcommand{\data}[1]{\texttt{#1}}

\newcommand{\mathbbm}[1]{\text{\usefont{U}{bbm}{m}{n}#1}} 

\newcommand{\R}{\mathbb{R}}

\DeclareMathOperator*{\argmin}{arg\,min}

\newcommand{\at}[2]{#1^{(#2)}}

\newcommand{\vbeta}{\boldsymbol{\beta}}
\newcommand{\vgamma}{\boldsymbol{\gamma}}
\newcommand{\braces}[1]{\left\{#1\right\}}
\newcommand{\paren}[1]{\left(#1\right)}
\newcommand{\brackets}[1]{\left[#1\right]}
\newcommand{\given}{\;\middle|\;}
\newcommand{\mean}[1]{\langle #1 \rangle }

\newcommand{\abs}[1]{\left\lvert #1 \right\rvert}
\newcommand{\one}[1]{\mathbbm{1}\!\!\brackets{#1}}

\newcommand{\model}{HCM\xspace}
\newcommand{\Model}{Hyperedge Copy Model\xspace}

\newcommand{\extantnodedist}{\boldsymbol{\gamma}}
\newcommand{\extantnodepar}{\gamma}
\newcommand{\novelnodedist}{\boldsymbol{\beta}}
\newcommand{\novelnodepar}{\beta}
\newcommand{\noderetentionprob}{\eta}
\newcommand{\params}{\boldsymbol{\theta}}
\newcommand{\simplex}{S}

\newcommand{\edgeset}{\mathcal{E}}
\newcommand{\nodeset}{\mathcal{N}}
\newcommand{\hypergraph}{\mathcal{H}}

\newcommand{\cN}[1][]{
    \ifthenelse{\equal {#1} {}}
        {\mathcal{N}}
        {\at{\mathcal{N}}{#1}}
}

\newcommand{\cE}[1][]{
    \ifthenelse{\equal {#1} {}}
        {\mathcal{E}}
        {\at{\mathcal{E}}{#1}}
}

\newcommand{\cF}[1][]{
    \ifthenelse{\equal {#1} {}}
        {\mathcal{F}}
        {\at{\mathcal{F}}{#1}}
}
\newcommand{\cH}[1][]{
    \ifthenelse{\equal {#1} {}}
        {\mathcal{H}}
        {\at{\mathcal{H}}{#1}}
}

\newcommand{\prob}[2][]{
    \ifthenelse{\equal {#1} {}}
    {
        \ifthenelse{\equal {#2} {}}
        {\mathbb{P}}
        {\mathbb{P}\paren{#2}}
    }
    {
        \ifthenelse{\equal {#2} {}}
        {\at{\mathbb{P}}{#1}}
        {\at{\mathbb{P}}{#1}\paren{#2}}
    }
}

\newcommand{\E}[2][]{
    \ifthenelse{\equal {#1} {}}
    {
        \ifthenelse{\equal {#2} {}}
        {\mathbb{E}}
        {\mathbb{E}\brackets{#2}}
    }
    {
        \ifthenelse{\equal {#2} {}}
        {\at{\mathbb{E}}{#1}}
        {\at{\mathbb{E}}{#1}\brackets{#2}}
    }
}

\newcommand{\empirical}[2][]{
    \ifthenelse{\equal {#1} {}}
        {\rho\brackets{#2}}
        {
            \ifthenelse{\equal {#2} {}}
            {\at{\rho}{#1}}
            {\at{\rho}{#1}\brackets{#2}}
        }
}

\newtheorem*{thm*}{Theorem}

\newtheorem*{conj*}{Conjecture}
\newtheorem*{clm*}{Claim}
\theoremstyle{definition}

\newtheorem*{dfn*}{Definition}

\newtheorem*{ex*}{Example}

\Crefname{dfn}{Definition}{Definitions}
\crefname{dfn}{definition}{definitions}

\newcommand{\eqdef}{\triangleq}

\usepackage{algorithm}
\usepackage[noend]{algpseudocode}
\usepackage{booktabs}
\usepackage[margin=1in]{geometry}

\hypersetup{
    hidelinks=true
}

\Crefformat{appendix}{Supplementary Section #2#1#3}

\begin{document} 

\title{Edge Correlations and Link Prediction in Growing Hypergraphs}
\date{\today}

\author{Xie He}
\thanks{Equal contributions}
\affiliation{Microsoft Research}

\author{Philip S. Chodrow}
\thanks{Equal contributions}
\affiliation{Department of Computer Science, Middlebury College}
\email{Corresponding author: pchodrow@middlebury.edu}

\author{Peter J. Mucha}
\affiliation{Department of Mathematics, Dartmouth College}

\begin{abstract}
    We propose a generative, mechanistic model of temporally-evolving hypergraphs in which hyperedges form via noisy copying of previous hyperedges. 
Our proposed model reproduces several stylized facts from many empirical hypergraphs, is learnable from data, and defines a likelihood over a complete hypergraph rather than ego-based or other sub-hypergraphs. 
Analyzing our model, we derive asymptotic descriptions of the node degree, edge size, and edge intersection size distributions in terms of the model parameters. 
We also show several features of empirical hypergraphs which are and are not successfully captured by our model. 
We provide a scalable stochastic expectation maximization algorithm with which we can fit our model to hypergraph data sets with millions of nodes and edges. 
Finally, we assess our model on a hypergraph link prediction task, finding that an instantiation of our model with just 11 parameters can achieve competitive predictive performance with large neural networks. 

\end{abstract}

\maketitle

\section{Introduction}
Many complex systems are composed of simple components participating in multi-way interactions. 
Such systems --- often called \emph{higher-order networks} \cite{bickWhatAreHigherOrder2023a} to distinguish them from networks composed of only pairwise interactions --- have received much attention in recent years. 
Higher-order networks preserve richer structural information about a system and therefore admit a broader array of measures, dynamics, and algorithms than their pairwise counterparts \cite{battistonNetworksPairwiseInteractions2020,battiston2021physics,pmlr-v97-chitra19a}. 
Higher-order networks are frequently represented as either hypergraphs or simplicial complexes \cite{bacciniWeightedSimplicialComplexes2022}. 
Whether a given system demands a higher-order representation, and which one to choose, are subtle modeling questions \cite{torresWhyHowWhen2021} driven by both the data analysis techniques to be used and the structure of the data itself. 

Hypergraphs are among the most flexible data structures for representing higher-order networks. 
In hypergraphs, each multi-way interaction is represented by an edge consisting of a set of nodes of arbitrary finite size. 
While simplicial complex representations require that every subset of every edge is also present in the data, hypergraphs require no such stipulation. 
Empirically, many hypergraphs come close to satisfying this subset inclusion criterion, but many others do not \cite{landrySimplicialityHigherorderNetworks2024}. 
While many hypergraph data sets include only the memberships of nodes in edges, other data sets may include node attributes \cite{badalyanHypergraphsNodeAttributes2023}, directedness or generalized node-edge roles \cite{gallo1993directed,chodrowAnnotatedHypergraphsModels2020}, and temporal information about the arrival or duration of interactions \cite{leeTHyMeTemporalHypergraph2021,myersTopologicalAnalysisTemporal2023}. 
The latter case is often described under the heading of \emph{temporal hypergraphs}, and is our modeling focus in this paper. 
Some examples of systems naturally represented by temporal hypergraphs include group socializing, email communication, and scholarly collaboration \cite{neuhauserConsensusDynamicsTemporal2021,sahasrabuddheModellingNonlinearConsensus2021,cencettiTemporalPropertiesHigherorder2021}.

Models of temporal hypergraphs can both shed light on the mechanisms underlying the evolution of higher-order networks and enable the prediction of future interactions.
Our modeling framework is motivated by a simple, fundamental difference between hypergraphs and dyadic graphs. 
In dyadic graphs, all edges contain two nodes (ignoring self-loops). 
Any pair of edges can therefore intersect on zero, one, or two nodes. 
The number of two-edge motifs \cite{miloNetworkMotifsSimple2002} in undirected graphs (in which self-loops are disallowed but multiedges are allowed) is therefore three. 
In contrast, in hypergraphs, an edge of size $i$ may intersect with another edge of size $j$ on a set of any size $k \leq \min\{i,j\}$. 
The number of possible two-edge motifs in a hypergraph \cite{leeHypergraphMotifsConcepts2020,lotitoHigherorderMotifAnalysis2022} with maximum edge-size $\bar{k}$ is therefore $O(\bar{k}^3)$, since the sizes of two edges, as well as the size of the intersection can all be distinguished. 
A wide range of intersection behavior is observed in empirical hypergraph data \cite{leeHowHyperedgesOverlap2021}, at rates much higher than explained by simple null models \cite{chodrowConfigurationModelsRandom2020}.

Several generative models have been proposed which produce large hypergraph edge intersections \cite{leeSurveyHypergraphMining2024}. 
The model of Lee, Choe, and Shin \cite{leeHowHyperedgesOverlap2021} reproduces several empirical intersection patterns in a hypergraph with specified edge-size and node degree distributions. 
Because these features of the data are specified in advance, the model is statistically generative but does not model mechanistic evolution in growing hypergraphs. 
The Correlated Repeated Unions (CRU) model of Benson, Kumar, and Tomkins \cite{bensonSequencesSets2018} is an explicitly mechanistic and temporal hypergraph model: arriving edges are formed as unions of noisy copies of previous edges. 
Copy models have a rich history in network science, serving, for example, as an early alternative to preferential attachment as a mechanistic explanation for heavy tails in the degree sequences of dyadic graphs \cite{kleinbergWebGraphMeasurements1999,soleModelLargeScaleProteome2002,vazquezModelingProteinInteraction2003,newman2018networks}. 
The CRU model is designed to generate hypergraphs in which the same subsets of nodes appear in multiple edges. 
This model, however, is designed for subsampled hypergraphs which may be viewed as single ``sequences of sets.''
Applying the model to a full hypergraph dataset requires the researcher to subsample the data into one or more such sequences; 
this limits the model's ability to describe hypergraphs in which seemingly disparate sequences may merge, and also prevents the researcher from evaluating a model likelihood on the complete data. 
Roh et al.~\cite{rohGrowingHypergraphsPreferential2023} details the degree distribution and edge-size distribution for a model of growing hypergraphs with preferential linking in which the evolution of the hypergraph is based entirely on adding new nodes to current edges and nodes. 
Our proposed model is perhaps most similar to the hypergraph preferential attachment models proposed by Avin et al.~\cite{avinRandomPreferentialAttachment2019} and Giroire et al.~\cite{giroirePreferentialAttachmentHypergraph2022} involve several types of updates to the hypergraph, such as isolated vertex addition and multiple types of edge addition. 
In each timestep, one of these actions is selected and performed. 
In contrast, our model involves a single, multipart update step. 
In each timestep, a new edge is formed as a noisy copy of an existing edge that was formed at a prior time, supplemented with nodes existing elsewhere in the hypergraph and novel nodes added after copying. 

One way to validate a model of hypergraph growth is to show that it reproduces macroscopic structural features observed in empirical data. 
We further validate our model via hyperedge link prediction: given observations of the hypergraph up to a specified time, we aim to predict which possible new edges are likely to form in the future. 
The link prediction task was first popularized for dyadic graphs \cite{liben-nowellLinkpredictionProblemSocial2007}. 
The generalization of link prediction to hypergraphs was popularized by Benson et al.~\cite{bensonSimplicialClosureHigherorder2018} and has since received treatment from a wide range of approaches \cite{lizotteHypergraphReconstructionUncertain2023,youngHypergraphReconstructionNetwork2021,chen2023survey}, with techniques including linear discriminative models \cite{bensonSimplicialClosureHigherorder2018}, neural discriminative models \cite{yadatiNHPNeuralHypergraph2020}, statistical generative models \cite{ruggeriCommunityDetectionLarge2023}, and mechanistic generative models \cite{bensonSequencesSets2018,avinRandomPreferentialAttachment2019,giroirePreferentialAttachmentHypergraph2022,giroirePreferentialAttachmentHypergraph2022a}. 
Link prediction in hypergraphs has a broad range of applications.  
In collaboration networks, predicting future collaborations among multiple entities can aid in resource allocation and project planning \cite{zhangHypergraphModelSocial2010}. 
In chemical networks, forecasting interactions among a set of molecules can contribute to drug discovery and understanding molecular processes \cite{tavakoliRxnHypergraphHypergraph2022}. 
In metabolic networks, predicting reactants and products can facilitate the discovery of novel metabolic pathways \cite{chen2023survey}. 
In logistics and supply chain management, predicting future connections in a hypergraph can optimize the flow of goods and resources \cite{suoExploringEvolutionaryMechanism2018}.
In social networks predicting future interactions can validate mechanistic models of complex human social behavior \cite{meng2024link}. 
In knowledge hypergraphs, predicting unseen multiway relations can help improve reasoning \cite{chen2022explainable}.

Our work is organized as follows.
In \Cref{sec:methods}, we describe the model update and derive asymptotic descriptions of the edge-size distribution, node degree distribution, and edge intersection rates. 
The generative nature of this \model also provides a principled statistical framework for fitting and evaluating model fit to data. 
However, because we do not observe the identity of the edge which is (noisily) copied in each timestep, direct optimization of the model likelihood is intractable. 
Instead, we develop a stochastic expectation-maximization algorithm \cite{cappeOnLineExpectationMaximization2009} for the inference task. 
In \Cref{sec:results}, we use stochastic expectation-maximization to fit our model to 27 empirical hypergraphs of sizes spanning several orders of magnitude.
We also evaluate the fitted model on a hyperedge prediction task with real-world hypergraphs, finding competitive predictive performance benchmarking against neural network methods \cite{yangLHPLogicalHypergraph2023, yadatiNHPNeuralHypergraph2020}. despite the extremely low-dimensional parameter space of the model. 
We close with discussion and suggestions for model generalizations in \Cref{sec:discussion}.

\section{Methods} \label{sec:methods}

\subsection{\Model (\model)} \label{sec:model}

\begin{figure*}[!ht]

\subfloat[\label{toy-model:a}]{%
  \includegraphics[width=0.6\columnwidth]{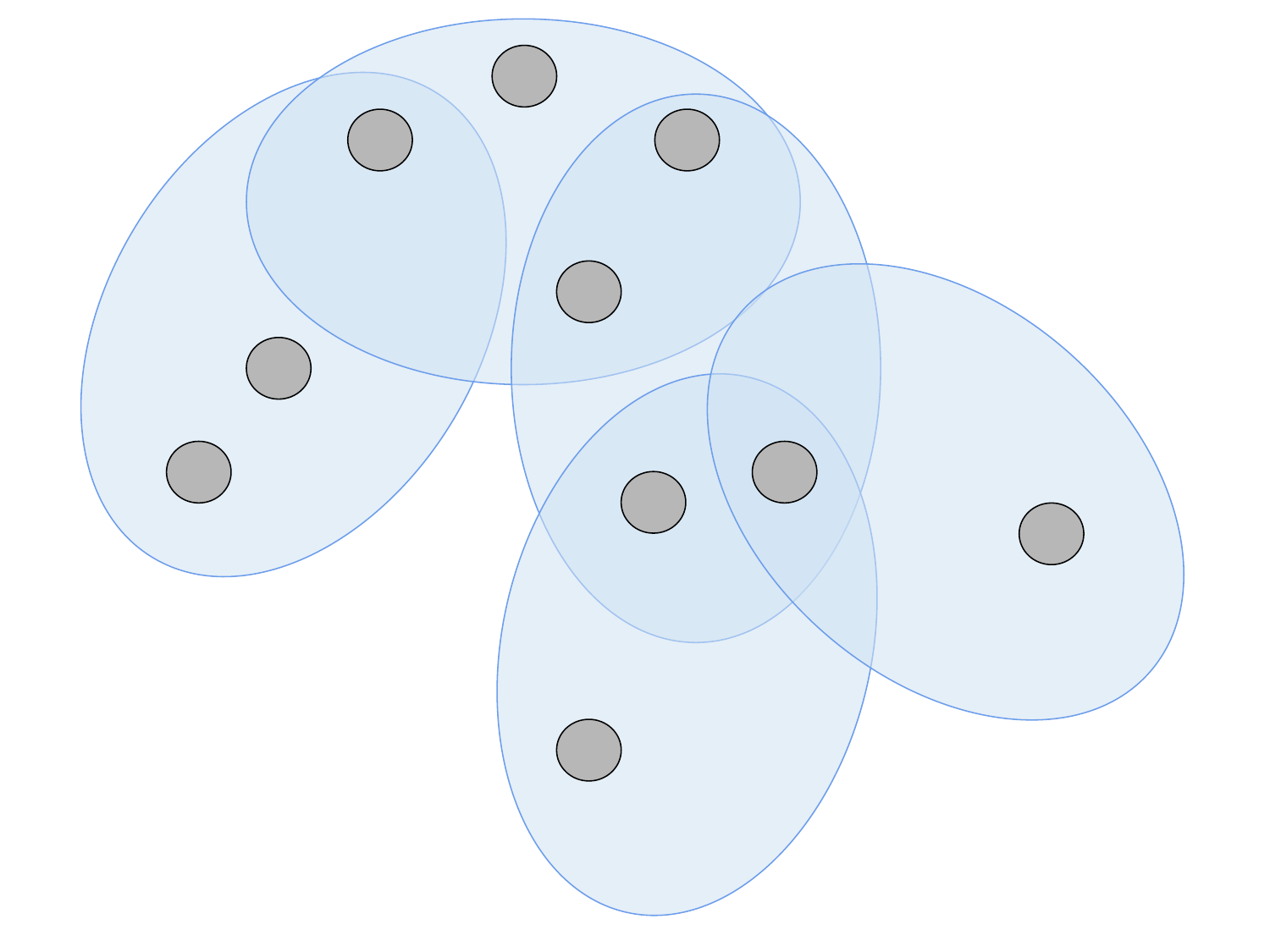}%
}\hfill
\subfloat[\label{toy-model:b}]{%
  \includegraphics[width=0.6\columnwidth]{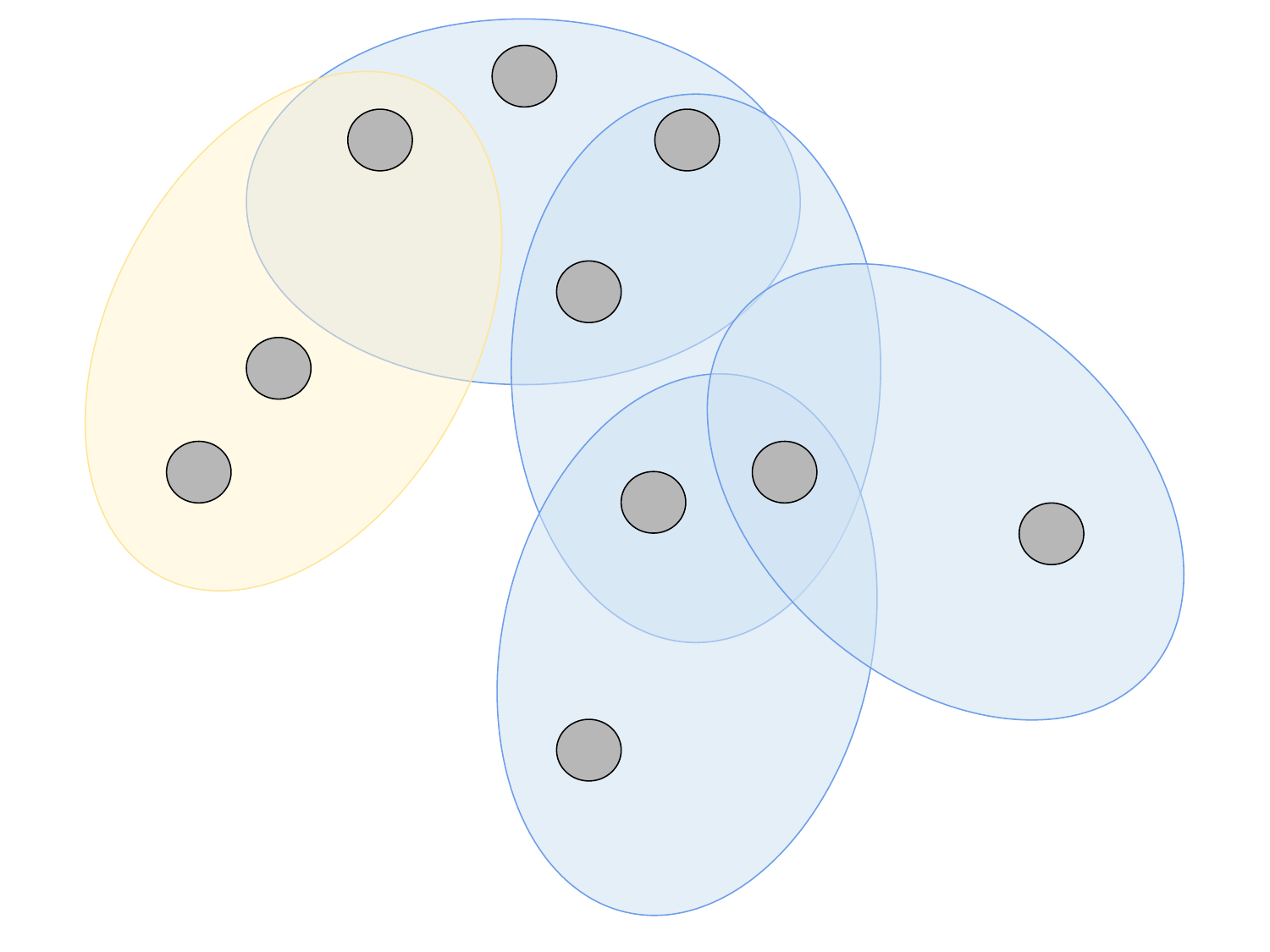}%
}\hfill
\subfloat[\label{toy-model:c}]{%
  \includegraphics[width=0.6\columnwidth]{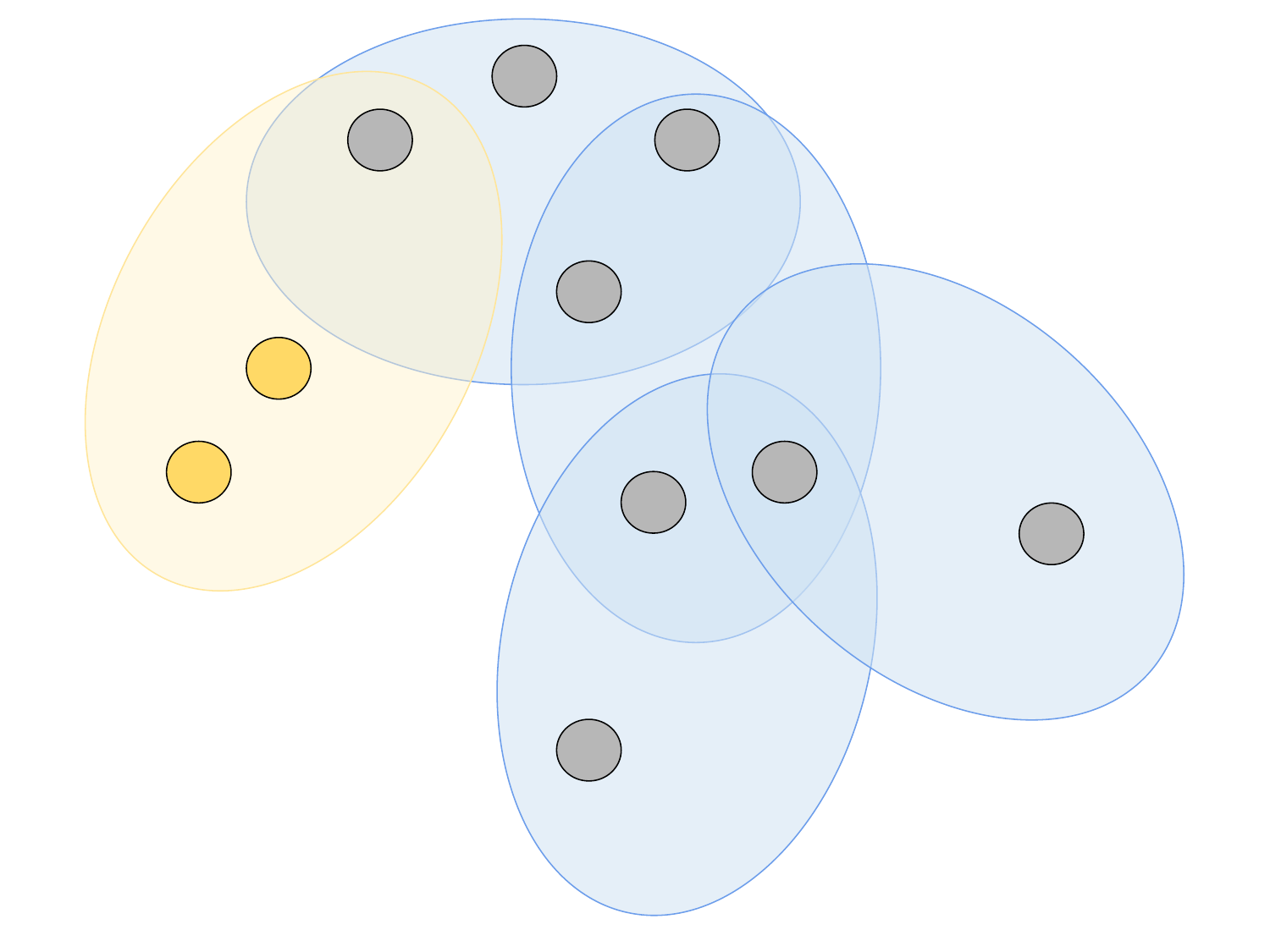}%
}

\subfloat[\label{toy-model:d}]{%
  \includegraphics[width=0.6\columnwidth]{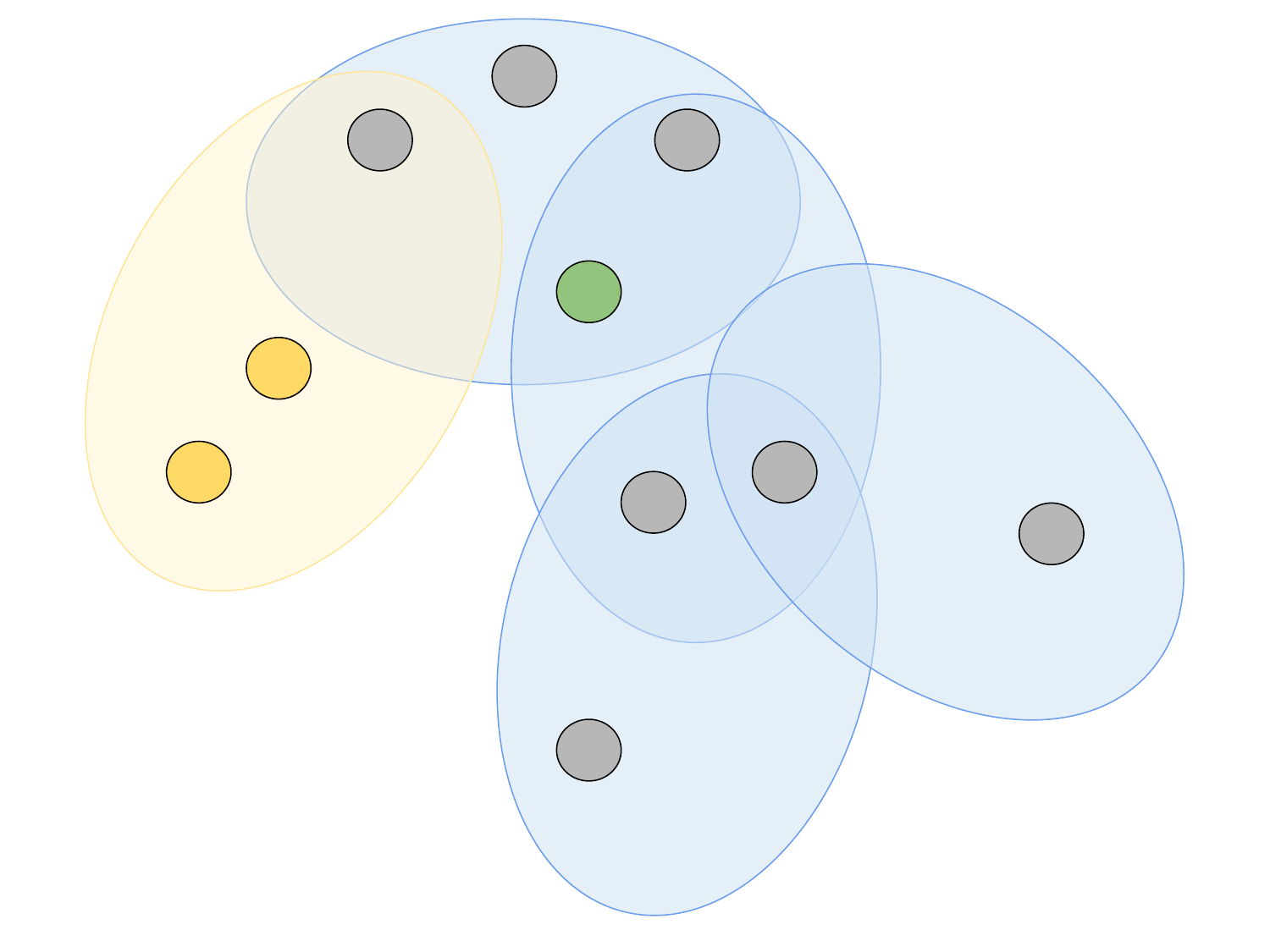}%
}\hfill
\subfloat[\label{toy-model:e}]{%
  \includegraphics[width=0.6\columnwidth]{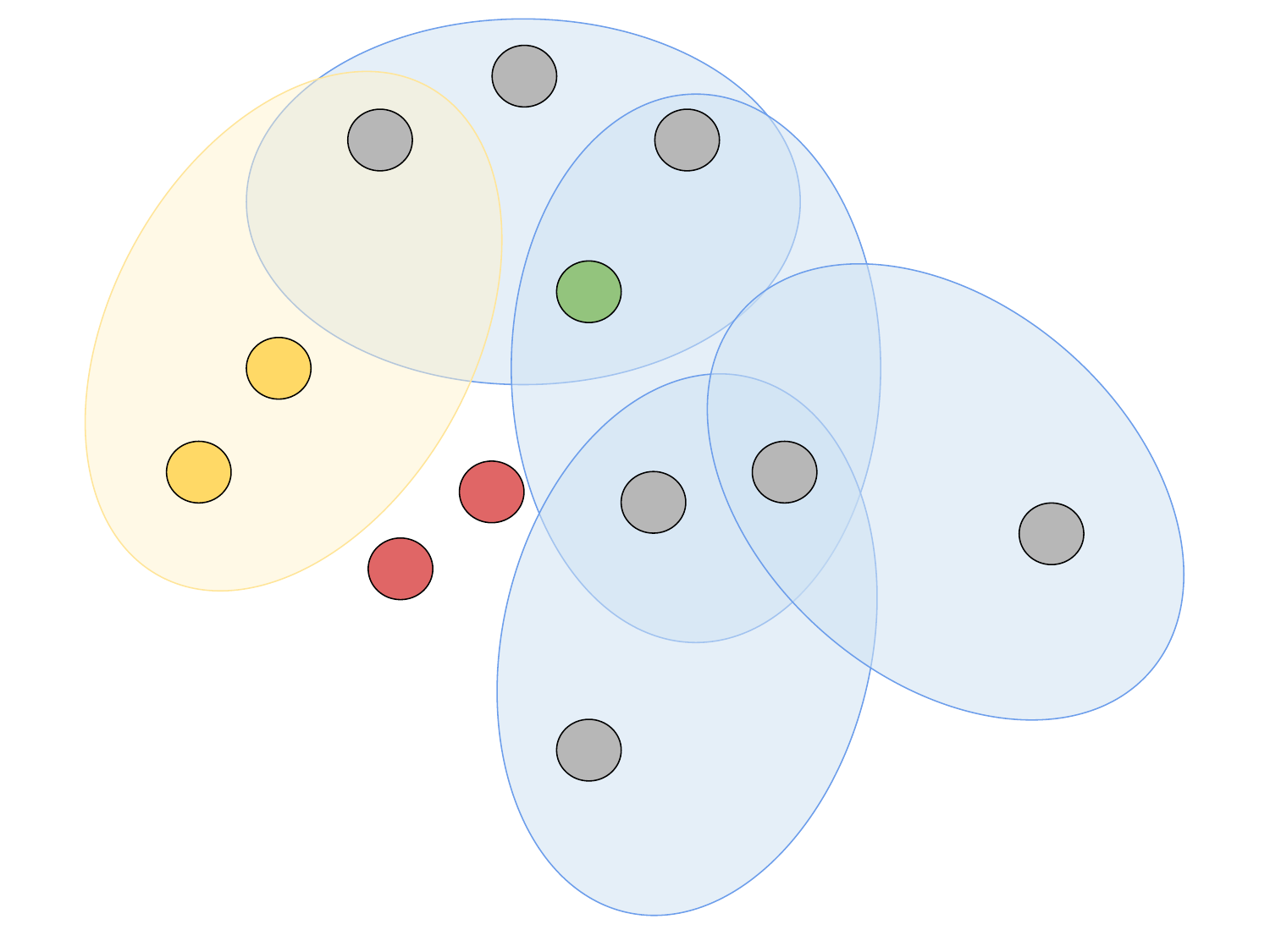}%
}\hfill
\subfloat[\label{toy-model:f}]{%
  \includegraphics[width=0.6\columnwidth]{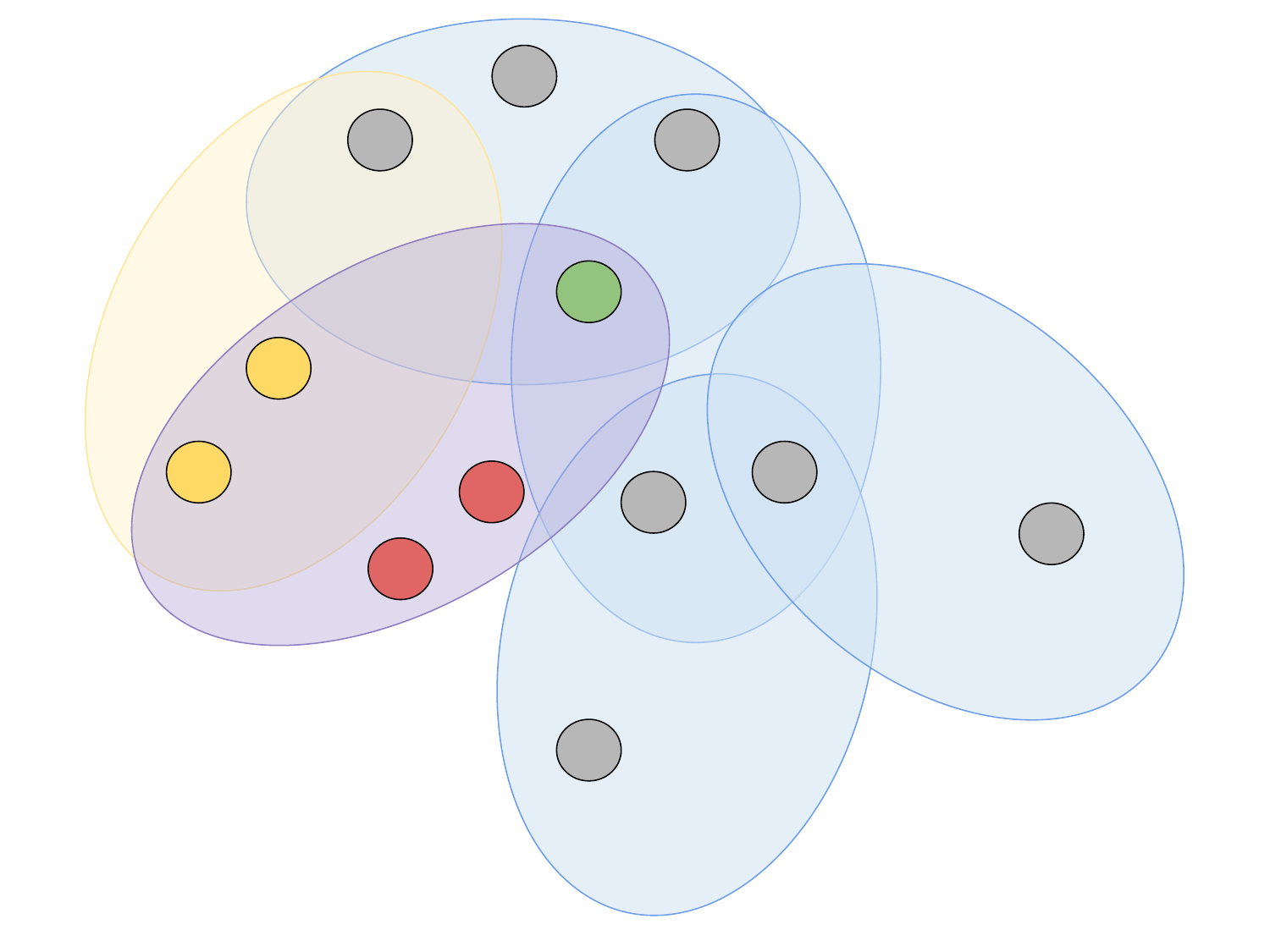}%
}

\caption{Schematic illustration of the edge generation process for our \Model (\model). (a): Current hypergraph $\at{\cH}{t}$. (b): \textbf{Edge selection}: An edge $f$ is selected uniformly at random (u.a.r.)\ from $\at{\edgeset}{t}$ and a node $v_0$ is selected u.a.r.\ from $f$ to seed the new edge $\at{e}{t+1}$. (c): \textbf{Edge sampling}: The remaining nodes $v \in f \setminus \braces{v_0}$ are each selected i.i.d.\ with probability $\eta$ for inclusion in $\at{e}{t+1}$. (d): \textbf{Extant node addition}: A number $g \sim \extantnodedist$ is sampled. Then, a set of $g$ distinct nodes is sampled u.a.r.\ from $\at{\nodeset}{t} \setminus f$ and added to $\at{e}{t+1}$. (e): \textbf{Novel node addition}: A number $b \sim \novelnodedist$ is sampled. Then, a set of $b$ distinct nodes is created and added to both $\at{\nodeset}{t+1}$ and $\at{e}{t+1}$. (f): Edge $\at{e}{t+1}$ is added to $\edgeset^{(t+1)}$.}
\label{fig:toy-model}
\end{figure*}

Our proposed model generates a sequence of growing hypergraphs. 
At each discrete timestep $t$, let $\at{\hypergraph}{t} = (\at{\nodeset}{t}, \at{\edgeset}{t})$ be a hypergraph with node set $\at{\nodeset}{t}$ and edge set $\at{\edgeset}{t}$.
Each hyperedge $e \in \at{\edgeset}{t}$ is a named set of nodes $e \subseteq \at{\nodeset}{t}$.
Multiedges --- in which two distinctly named edges are equal as sets --- are permitted. 
The degree of node $v$ in hypergraph $\at{\hypergraph}{t}$ is $\at{d}{t}_v = \sum_{e \in \at{\edgeset}{t}} \one{v \in e}$, the number of edges containing $v$ in $\at{\edgeset}{t}$.

We consider only a single method of updating the hypergraph from time $t$ to time $t+1$, which consists of the following four substeps: 
\begin{enumerate}
    \item \textbf{Edge selection}: a single edge $f$ is selected uniformly at random (u.a.r.) from  $\at{\edgeset}{t}$.
    Then, one node $v_0$ is selected u.a.r.\ from $f$ to seed the new edge $\at{e}{t+1}$.
    \item \textbf{Edge sampling}: the remaining nodes $v \in f \setminus \braces{v_0}$ are each included in $\at{e}{t+1}$ independently with probability $\eta \in [0,1]$.
    \item \textbf{Extant node addition}: a nonnegative integer $g$ is sampled from a distribution which we express as a probability vector $\extantnodedist$. 
    We require that $\extantnodedist$ be an element of $\simplex^{\bar{k}}$, where $\simplex^{\bar{k}}$ is the set of probability vectors indexed $0$ through $\bar{k}$.
    The choice of $\bar{k}$ gives the largest possible number of extant nodes which can be added in a single update step.
    With a slight abuse of notation, we write $g\sim \extantnodedist$ to denote this sampling process.  
    Then, $g$ distinct nodes are selected u.a.r.\ from $\at{\nodeset}{t}$ and added to $\at{e}{t+1}$.
    \item \textbf{Novel node addition}: a nonnegative integer $b \sim \novelnodedist$ is sampled, where again $\novelnodedist$ is a probability vector, and $b$ distinct nodes are created and added to $\at{e}{t+1}$.
\end{enumerate}
After forming $\at{e}{t+1}$, we define the updated hypergraph $\at{\hypergraph}{t+1} = (\at{\nodeset}{t} \cup \braces{\at{e}{t+1}}, \at{\edgeset}{t} \cup {\at{e}{t+1}})$. 
\Cref{alg:model} gives a formalized summary of the model, and \Cref{fig:toy-model} gives a schematic graphical illustration. 
The parameters of the model are the copy probability $\eta$, the extant node distribution $\extantnodedist$, and the novel node distribution $\novelnodedist$. 
For notational compactness, we will use the vector $\params = (\eta, \extantnodedist, \novelnodedist)$ to refer to the concatenation of these three parameters. 
The total number of scalar parameters of the model is $2\bar{k} + 1$.




\begin{algorithm}[H]
    \caption{\Model (\model) update step}\label{alg:model}
    \begin{algorithmic}
    \Require $\at{\hypergraph}{t} = \paren{\at{\nodeset}{t}, \at{\edgeset}{t}}$, $\eta \in [0,1]$, $\extantnodedist \in \simplex^{\bar{k}}$, $\novelnodedist \in \simplex^{\bar{k}}$
    \State $e \gets \emptyset$
    \State Sample $f \sim \mathrm{Uniform}\paren{\at{\edgeset}{t}}$ \Comment{Edge selection}
    \State Sample $v_0 \sim \mathrm{Uniform}\paren{f}$ \Comment{Edge sampling}
    \State $\at{e}{t+1} \gets e\cup \braces{v_0}$ 
    \For{node $v \in f\setminus v_0$}
    \State With probability $\eta$, $\at{e}{t+1} \gets \at{e}{t+1} \cup \braces{v}$   
    \EndFor
    \State Sample $g \sim \extantnodedist$ \Comment{Extant node addition}
    \State Sample $V \sim \mathrm{Uniform}\paren{\genfrac{}{}{0pt}{2}{\at{\nodeset}{t}\setminus f}{g}}$ 
    \State $\at{e}{t+1} \gets \at{e}{t+1} \cup V$
    \State Sample $b \sim \novelnodedist$  \Comment{Novel node addition}
    \State $n \gets \abs{\at{\nodeset}{t}}$
    \State $\nodeset^{(t+1)} \gets \at{\nodeset}{t} \cup \braces{v_{n+1}, \ldots, v_{n+b}}$ 
    \State $\at{e}{t+1} \gets \at{e}{t+1} \cup \braces{v_{n+1}, \ldots, v_{n+b}}$
    \State $\edgeset^{(t+1)} \gets \at{\edgeset}{t} \cup \at{e}{t+1}$
    \State \Return{$\at{\hypergraph}{t+1}=\paren{\at{\nodeset}{t+1}, \at{\edgeset}{t+1}}$}
    \end{algorithmic}
\end{algorithm}

\subsection{Asymptotic degree and edge-size distributions}

We now derive several asymptotic properties of our proposed \model. 
We first derive the asymptotic mean edge size, as well as a linear system describing the complete edge size distribution.  
Let $\mean{k}$ be the mean edge size in $\at{\hypergraph}{t}$.
We will compute this mean under an assumption of stationarity.
Each time an edge is constructed, there are in expectation $1 + \eta(\mean{k}-1)$ nodes sampled in the edge sampling step, $\mu_{\extantnodedist}$ nodes sampled in the extant node addition step, and $\mu_{\novelnodedist}$ nodes sampled in the novel node addition step, where $\mu_{\cdot}$ denotes the mean of the corresponding distribution. 
At stationarity, we therefore have the self-consistent equation 
\begin{align}
    \mean{k} = 1 + \eta(\mean{k}-1) + \mu_{\extantnodedist} + \mu_{\novelnodedist}\;, \label{eq:mean-edge-size-eq}
\end{align}
from which it follows that, provided $\eta < 1$, 
\begin{align}
    \mean{k} = \frac{1 - \eta + \mu_{\extantnodedist} + \mu_{\novelnodedist}}{1-\eta}\;. \label{eq:mean-edge-size}
\end{align}
When $\eta = 1$, the edge size is nonstationary and grows arbitrarily large, as reflected in the divergence of \cref{eq:mean-edge-size}. 

To describe the edge size distribution, let $\mathbf{W} \in \R^{\bar{k}\times \bar{k}}$ be the matrix whose entry $w_{ij}$ gives the probability that the produced edge $\at{e}{t+1}$ has size $i$ given that the sampled edge $f$ has size $j$. 
As we show in \Cref{sec:edge-sizes}, the entries of $\mathbf{W}$ have closed-form expressions in terms of the parameter vector $\params$: 
\begin{align}
    w_{ij} = \sum_{\ell = 0}^{j-1} \sum_{h = 0}^{i - \ell} \binom{j-1}{\ell} \eta^{\ell} (1-\eta)^{j - \ell - 1} \extantnodepar_h \novelnodepar_{i - \ell - h}\;. \label{eq:edge-size-distribution-matrix}
\end{align}
The stationary distribution of edge sizes under our model is then given by the Perron eigenvector of the matrix $\mathbf{W}$. 
We give several examples of computing the stationary distribution of edge sizes for synthetic and empirical data sets using \cref{eq:edge-size-distribution-matrix} in \Cref{fig:degree-edge-size}.

Turning now to the degree distribution, we first calculate the mean degree $\mean{d}$. 
In any hypergraph of $n$ nodes and $m$ edges, it holds that  ${n}\mean{d} = {m \mean{k}}$. 
At large $t$, in expectation $m/n \approx {1}/{\mu_{\novelnodedist}}$, since on average there are $\mu_{\novelnodedist}$ nodes added per new edge. 
Provided that $\eta <1$ (so that $\mean{k}$ is finite), we therefore have 
\begin{align}
    \mean{d} = \frac{\mean{k}}{\mu_{\novelnodedist}} = \frac{1 - \eta + \mu_{\extantnodedist} + \mu_{\novelnodedist}}{\mu_{\novelnodedist}(1-\eta)}\;. \label{eq:mean-degree}
\end{align}

Furthermore, the tails of the degree distribution are power-law with an exponent that depends on the model parameters. 
In \Cref{sec:power-law-tails} we follow a derivation by Mitzenmacher \cite{mitzenmacherBriefHistoryGenerative2004} to argue that, for large $d$, the proportion $p_d$ of nodes with degree $d$ is approximated, for large $d$, by the power law $p_d \propto d^{-\zeta}$, where the exponent $\zeta$ is given by 
\begin{align}
    \zeta = 1 + \frac{1 - \eta + \mu_{\extantnodedist} + \mu_{\novelnodedist}}{1 - \eta(1 - \mu_{\extantnodedist} - \mu_{\novelnodedist})}\;. \label{eq:power-law-exponent}
\end{align}
An intuitive explanation for the occurrence of a power law in this context is that the edge selection and sampling steps of the \model choose nodes in proportion to the number of edges in which those nodes are present; i.e.\ their degree. 
Degree-proportional sampling is a classical mechanism for generating power-law degree distributions \cite{barabasiEmergenceScalingRandom1999,kleinbergWebGraphMeasurements1999}. 
We show examples of the power law exponent predicted by our model in comparison to synthetic and empirical data sets in  \Cref{fig:degree-edge-size}.

\subsection{Asymptotic edge intersection sizes}

\begin{figure}[!ht]
    \includegraphics[width=1\linewidth]{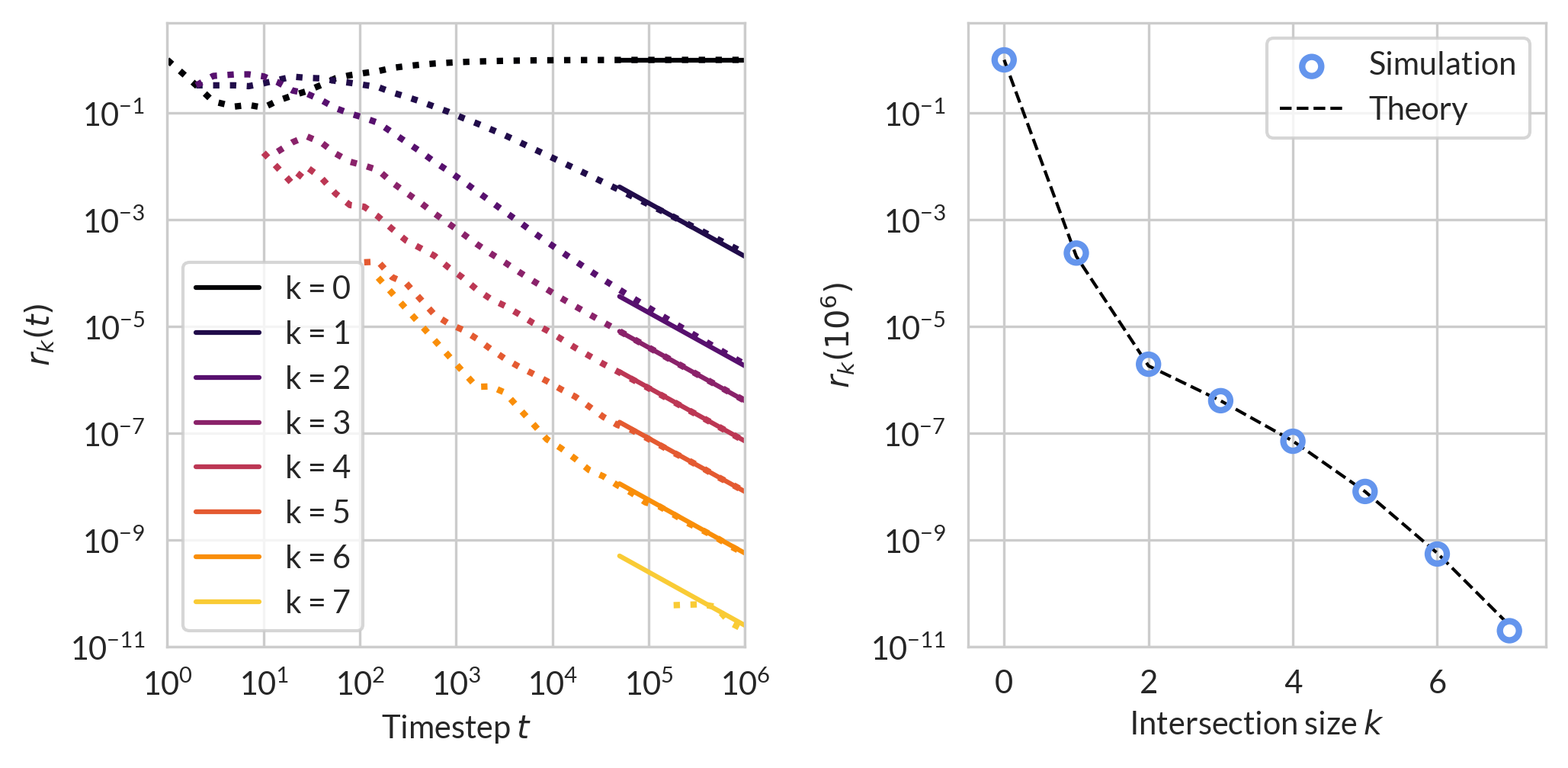}
    \caption{Illustration of the edge intersection asymptotics given by \cref{eq:intersection-asymptotics}. 
    We simulated an \model for $10^6$ timesteps with parameters $\eta = 0.3$, $\vbeta$ uniform on $\braces{1,2}$, and $\vgamma$ uniform on $\braces{0,1}$.
    (Left): Scaling of intersection size densities $r_k$ as a function of timestep $t$. 
    Dotted lines give empirical intersection rates. 
    Solid lines for $k \geq 1$ give the predicted scaling $m^{-1} \sum_{ij}q_{ijk}$, with $q_{ijk}$ computed as the entries of the leading eigenvector for a matrix $\mathbf{C}$ determined by the model parameters.
    Note the log-log axes. 
    (Right): Proportion $r_k$ of pairs of edges intersecting on a set of size $k$ at the end of our simulation after $10^6$ timesteps, compared with predictions obtained from \cref{eq:intersection-asymptotics}. 
    } 
    \label{fig:intersection-asymptotics}
\end{figure}

It is possible to compute the exact asymptotic structure of the densities of pairwise intersections in our proposed model. 
We express this structure in terms of the joint distribution 
\begin{align}
    r_{ijk} \eqdef \rho \paren{\abs{e} = i, \abs{f} = j, \abs{e\cap f} = k \given | e\succ f}\;, 
\end{align}
where $e\succ f$ indicates that edge $e$ appears later than edge $f$.
We present a probabilistic argument in \Cref{sec:intersection-sizes} for a claim describing the asymptotic structure of the intersection sizes in our model.
Under this claim, if $\beta_0 < 1$, there exist constants $q_{ijk}$ such that, as the number of edges $m$ grows large: 
\begin{itemize}
    \item The intersection sizes $r_{ijk}$ are related to the constants $q_{ijk}$ as 
    \begin{align}
        r_{ijk} = \begin{cases}
            q_{ijk} + O \paren{m^{-1}} &\quad k = 0 \\
            m^{-1}q_{ijk} + O\paren{m^{-2}} &\quad \text{otherwise\;.}
        \end{cases} \label{eq:intersection-asymptotics}
    \end{align}
    \item The constants $q_{ijk}$ can be computed as the entries of the unique nonnegative eigenvector of a matrix $\mathbf{C}$ determined by the parameters $\eta$, $\extantnodedist$, and $\novelnodedist$. 
\end{itemize}
These asymptotics are illustrated in \Cref{fig:intersection-asymptotics} on a large synthetic hypergraph simulated according to the \model{}. 
We show the proportion $r_{k} = \sum_{ij}r_{ijk}$ of pairs of edges of any size intersecting on a set of size $k$ over time (left) and at the final timestep (right), and compare these to the asymptotic predictions of \cref{eq:intersection-asymptotics}.
The predictions agree closely with simulated values, matching both the $m^{-1}$ scaling and the predicted intercepts. 

\subsection{Other asymptotic properties}

\begin{figure*}[!ht]
    
    \includegraphics[width=1\linewidth]{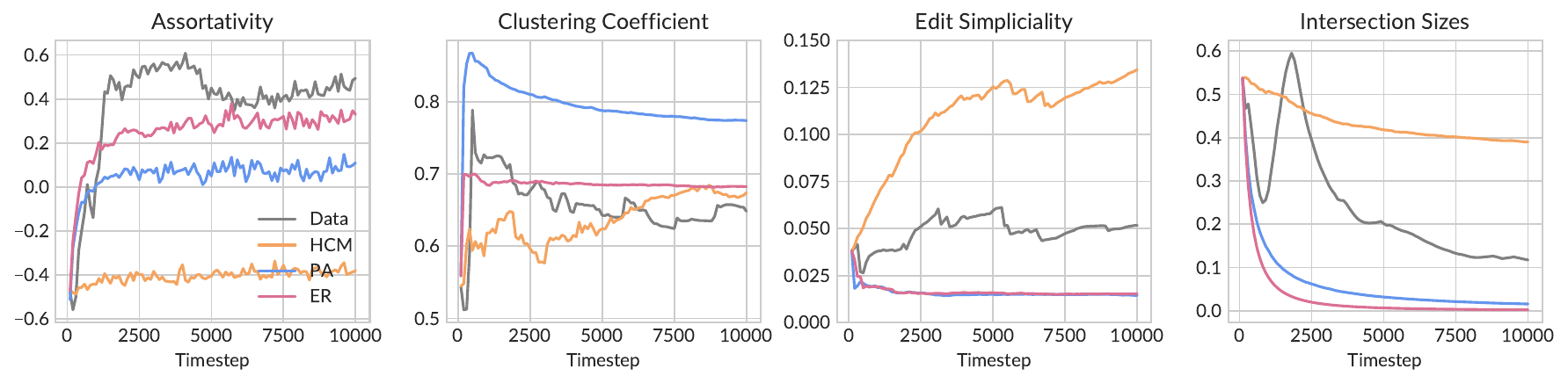} 
    \caption{
    Structural properties of the \data{email-enron} temporal hypergraph and three synthetic models. We note that assortativity is highly dataset-dependent, and the behavior of different models changes drastically with different datasets. Specifically, for edge intersection size, we observe that only \model has demonstrated non-diminishing intersection sizes over time steps, as shown in the same experiment on several other example datasets in \Cref{fig:additionalM4}.
        }
    \label{fig:four-properties}
\end{figure*}

Our \model displays several other asymptotic structural properties which differ simpler models of evolving hypergraphs (\Cref{fig:four-properties}). 
We fit the \model to the \data{email-enron} dataset using the stochastic expectation-maximization (SEM) algorithm described in the following section. 
We simulated a synthetic hypergraph generated according to the model. 
We also simulated two other synthetic hypergraphs: one generated according to a temporal Erd\H{o}s-R\'enyi-type model (ER) that replicates the edge-size distribution of the fitted \model, as well as a preferential attachment-type model (PA) which also replicates the power law degree distribution.
Details on these two alternative models can be found in \cref{sec:alternative-models}.  
We measured four structural properties of the empirical \data{email-enron} hypergraph and the three synthetic hypergraphs: uniform degree assortativity \cite{chodrowConfigurationModelsRandom2020}, clustering coefficient \cite{gallagherClusteringCoefficientsProtein2013}, edit simpliciality \cite{landrySimplicialityHigherorderNetworks2024}, and edge intersections \cite{landrySimplicialityHigherorderNetworks2024} (values shown in \autoref{fig:four-properties}). 

In the case of \data{email-enron}, we observe that the behavior of the \model is quite distinct from that of the ER and PA models. 
The \model replicates the degree assortativity less well than either the ER or PA models and replicates the clustering coefficient roughly as well as the ER model. 
For edit simpliciality and intersection sizes, the \model does not quantitatively capture the correct measurement, but does qualitatively express that these quantities are not asymptotically vanishing, in contrast to the predictions by both the ER and PA models. 
We show the same experiment on several other datasets in  \Cref{fig:additionalM4}.

\subsection{Inference via Stochastic Expectation Maximization (SEM)} \label{sec:inference}

We use maximum-likelihood estimation to learn the parameters $\noderetentionprob$, $\extantnodedist$, and $\novelnodedist$ of our proposed \model from a dataset containing time-stamped hyperedges. 
The \model has the structure of a latent-variable model: the edge $\at{e}{t+1}$ that was added in timestep $t+1$ is observed, but the edge $f$ that was sampled in the edge-selection step to generate $\at{e}{t+1}$ is not. 
This structure lends itself naturally to optimization via the expectation-maximization (EM) algorithm \cite{dempster1977maximum}.
Let $\at{F}{t+1}$ be the true but unobserved edge that was sampled in the edge-selection step. 
Given the observation of $\at{e}{t+1}$ and some current estimate $\params$ of the parameters, the probability that the true $\at{F}{t+1}$ was edge $f$ is
\begin{widetext}
\begin{align*}
    \prob{\at{F}{t+1} = f \given \at{e}{t+1} ; \params} &= \frac{\prob{\at{e}{t+1}, \at{F}{t+1} = f; \params}}{\prob{\at{e}{t+1}; \params}} \\ 
    &= \frac{\prob{\at{e}{t+1} , \at{F}{t+1} = f; \params}}{\sum_{f'}\prob{\at{e}{t+1} ,  \at{F}{t+1} = f'; \params}}\; \\ 
    &= \frac{\prob{\at{e}{t+1} \given \at{F}{t+1} = f; \params}}{\sum_{f'}\prob{\at{e}{t+1} \given  \at{F}{t+1} = f'; \params}}\;,
\end{align*}
\end{widetext}
where the final line follows because the selection of $\at{F}{t+1} = f$ in the HCM is uniform over the edge set $\at{\edgeset}{t}$.
The probabilities appearing in this final expression are determined by \Cref{alg:model} and can be computed in closed form. 
The general EM algorithm applied to the single observation of $\at{e}{t+1}$ proceeds by maximizing the expectation of the log-likelihood with respect to a new parameter estimate, with the expectation taken with respect to $\prob{\at{F}{t+1} = f \given \at{e}{t+1} ; \params}$, which we now abbreviate $\at{p}{t+1}(f; \params)$ for notational compactness. 
In our case, this maximization problem has a closed-form solution in terms of the vector $\mathbf{s}$ of expected sufficient statistics of the distributions involved in sampling. 
This vector has components 
\begin{align}
    s_1        &= \sum_{f}\at{p}{t+1}(f; \params) \abs{f\cap \at{e}{t+1}}\;, \label{eq:expected-ss-1}\\ 
    s_2        &= \sum_{f}\at{p}{t+1}(f; \params) \abs{f\setminus \at{e}{t+1}}\;, \label{eq:expected-ss-2}\\ 
    s_{3,\ell} &= \sum_{f}\at{p}{t+1}(f; \params) \one{\abs{\at{e}{t+1} \setminus \at{\nodeset}{t}} = \ell}\;. \label{eq:expected-ss-3}\\ 
    s_{4,\ell} &= \sum_{f}\at{p}{t+1}(f; \params) \one{\abs{\paren{\at{e}{t+1} \setminus f} \cap \at{\nodeset}{t}} = \ell} \;. \label{eq:expected-ss-4}
\end{align}
Once $\mathbf{s}$ is computed, the maximum likelihood estimates for the parameters are 
\begin{align}
    \hat{\noderetentionprob} = \frac{s_1-1}{s_1 + s_2-1}\;, \quad 
    \hat{\extantnodepar}_\ell = s_{3,\ell}\;, \quad 
    \hat{\novelnodepar}_\ell = s_{4,\ell}\;. \label{eq:m-step}
\end{align}
Importantly, we are guaranteed that $s_1\geq 1$  by the requirement of \Cref{alg:model} that $\abs{f\cap \at{e}{t+1}}\geq 1$ for all $f$ with $\at{p}{t+1}(f; \params)>0$.

In standard, full-batch EM, we would compute averages of the expected sufficient statistics across the entire sequence of new edges, and then use \cref{eq:m-step} to form new parameter updates. 
We would then repeat this process until convergence to a local maximum of the likelihood function, which is guaranteed by standard theory. 
For our setting, however, the full-batch EM algorithm is infeasible due to the computational cost of forming the distributions $\at{p}{t+1}(f; \params)$, 
requiring summation over all $t$ edges that arrived prior to time $t+1$. 
This sum thus includes order $m^2$ terms for \emph{each} timestep, where $m$ is the number of edges in the hypergraph, which  
is computationally prohibitive for hypergraphs with $m\gtrsim 10^4$ on modern personal computers.

We therefore instead consider a stochastic variant of EM \cite{cappeOnLineExpectationMaximization2009}. 
Starting from an arbitrary initial vector $\hat{\mathbf{s}}(0)$ of estimates of the expected sufficient statistics, we update with an exponentially moving average. 
In each algorithmic step $\tau$ of stochastic EM, we sample an edge $e$ uniformly at random. 
We then construct the vector $\mathbf{s}({\tau})$ of expected sufficient statistics from the observation of according to \cref{eq:expected-ss-1,eq:expected-ss-2,eq:expected-ss-3,eq:expected-ss-4}. 
Next, we update $\hat{\mathbf{s}}(\tau)$ according to
\begin{align}
    \hat{\mathbf{s}}(\tau+1) = \paren{1-\rho(\tau)}\hat{\mathbf{s}}(\tau) + \rho(\tau) \mathbf{s}(\tau). 
\end{align}
Here, $\rho(\tau)$ is a learning schedule that determines the rate of update in the estimate of the expected sufficient statistics.
The running estimate $\hat{\params}(\tau)$ of the parameter vector $\params$ is obtained from \cref{eq:m-step}, using $\mathbf{\hat{s}}(\tau)$ in place of $\mathbf{s}$. 
We use a learning schedule $\rho(\tau) = \tau^{-1}$ and terminate the algorithm when the relative change in the estimate of $\eta$ between the most recent 100-step windows falls below $10^{-2}$.

Although the above procedure is described for temporal hypergraphs, we also apply the same stochastic EM approach to non-temporal hypergraphs by imposing a randomized pseudo-temporal ordering. 
Specifically, we randomly assign edges to synthetic time steps and ensure that no edge is selected more than once in the sampling process, thereby mimicking the temporal structure assumed in the model. 
This allows us to perform likelihood-based inference by treating the randomized sequence as input to the same generative model. 
Although the temporal structure is synthetic, the parameters estimated in this manner still allow us to assign meaningful likelihood scores to observed edges, enabling tasks such as link prediction in non-temporal settings. We emphasize that while this randomized ordering permits parameter estimation and link scoring, metrics such as $\text{AUC}(t)$, which rely on true temporal dynamics, are only applicable for genuinely time-resolved data.

\section{Results} \label{sec:results}

\subsection{SEM-\model on empirical data}
We used SEM to fit our \model to 27 empirical hypergraphs provided by the XGI package for Python \cite{landryXGIPythonPackage2023}. 
Clock-times to convergence ranged from 1.6 seconds in the case of the \data{diseasome} dataset (516 nodes and 903 edges) to $2.9\times 10^4$ seconds (8 hours) for the \data{threads-stack-overflow} dataset ($2.7\times 10^6$ nodes and $1.1 \times 10^7$ edges). 
We give convergence times and more detailed descriptions of learned parameters in \Cref{tab:parameter-values}.  

\begin{figure*}
    \includegraphics[scale=0.55]{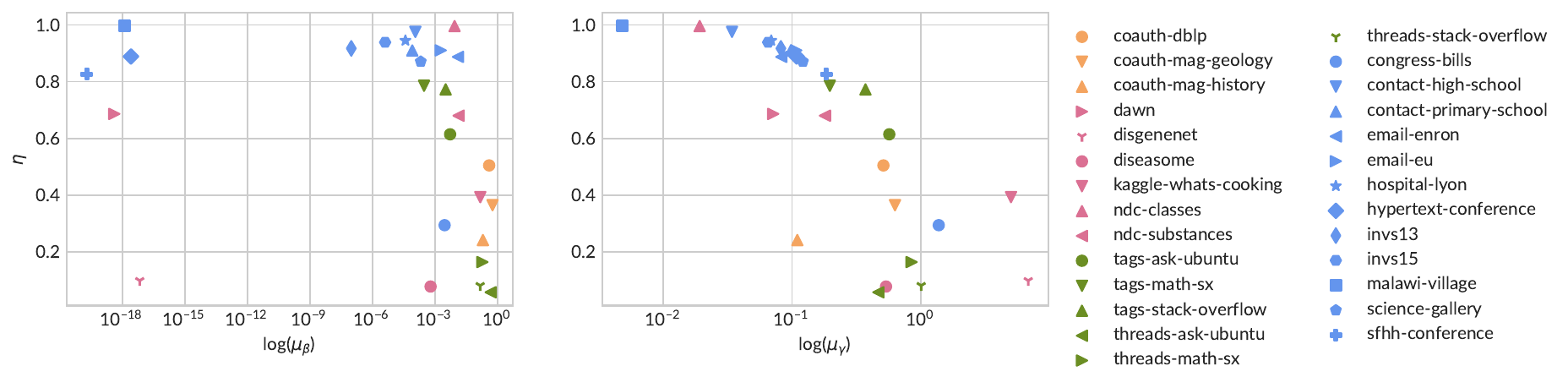}
    \caption{
    (Best viewed in color). Visual summary of parameters obtained by SEM fits of our \model to empirical hypergraphs. 
    We show $\eta$ and the expectations $\mu_{\vbeta} = \sum_{i} i\beta_i$ and $\mu_{\vgamma} = \sum_i i \gamma_i$.
    We use linear scale for $\eta$ and log scale for $\mu_{\vbeta}$ and $\mu_{\vgamma}$. 
    When fitting the model using SEM, The default length $\bar{k}$ of $\vbeta$ and $\vgamma$ is set to be equal to the largest edge size in the corresponding empirical hypergraphs. 
    Coauthorship datasets are shown in orange; biological datasets in pink; information datasets in green, and social interaction datasets in blue. 
    }
    \label{fig:parameters}
\end{figure*}

\Cref{fig:parameters} summarizes the parameters obtained by SEM fits of our \model to empirical hypergraphs.
We group the datasets into four broad categories: coauthorship, biological, information, and social interaction. 
The social interaction networks in our dataset have very high rates $\eta$ of edge-copying, along with relatively low rates of extant or novel node addition. 
Coauthorship networks tend to display lower rates of edge-copying and higher rates of extant and novel node addition. 
Biological and information networks display less clear patterns, with different networks in these categories displaying very different estimated parameter values. 
On the whole, most datasets exhibit seemingly small values of $\mu_{\extantnodedist}$, consistently below one, indicating low rates of adding extant nodes elsewhere in the hypergraph to the edge being copied. The \data{kaggle} graph, however, stands out as an outlier among the larger hypergraphs, introducing a substantial number of new nodes at each timestep.

\begin{figure*}[!ht]
    \includegraphics[width=1\linewidth]{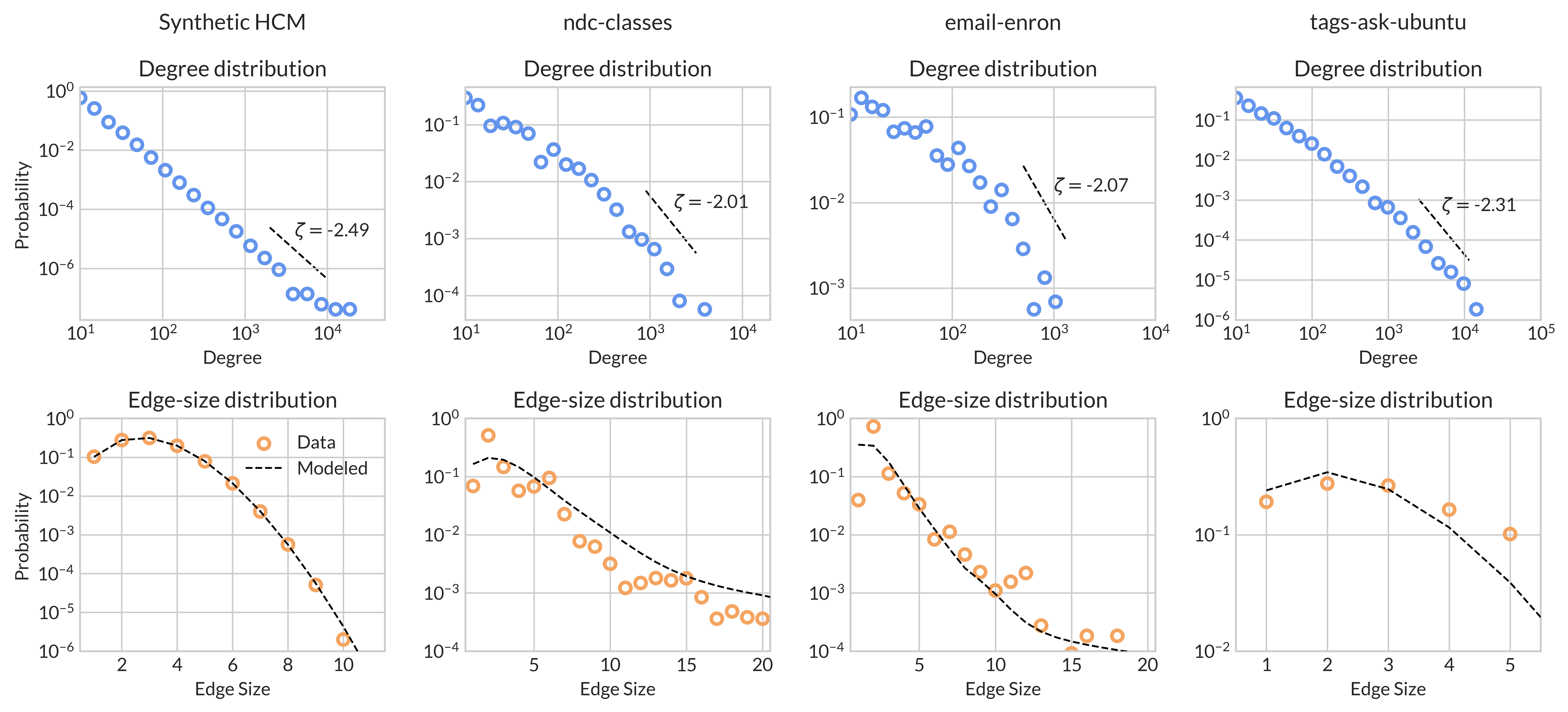}
    \caption{Degree distributions and edge-size distributions for one synthetic HCM with $10^6$ edges and three empirical datasets: \data{ndc-classes}, \data{email-enron}, and \data{tags-ask-ubuntu}.
    In the top row, dashed lines indicate the exponent of the power-law describing the asymptotic degree distribution using parameters inferred via SEM and \cref{eq:power-law-exponent}. 
    In the bottom row, the dashed curve gives the modeled asymptotic edge-size distribution using inferred parameters to compute the matrix $\mathbf{W}$ described by \cref{eq:edge-size-distribution-matrix} and its associated Perron eigenvector. 
    For the synthetic hypergraph (first column), the true parameters are used to construct the approximations and no inference is performed.
    The \data{ndc-classes} and \data{email-enron} edge-size distributions have been truncated for visualization purposes. 
    }
    \label{fig:degree-edge-size}
\end{figure*}

As one form of validation, we compare the actual degree and edge-size distributions of several datasets to the asymptotic descriptions provided by \cref{eq:edge-size-distribution-matrix,eq:power-law-exponent}, using the parameters obtained by SEM.
We show this comparison in \Cref{fig:degree-edge-size}. 
We show the slope corresponding to the power-law exponent for the degree-distribution of the \model as well as the complete modeled distribution of edge sizes.  
We find rough qualitative agreement in both cases, despite the small number of parameters that the \model uses to describe each dataset. 
This appears true despite the considerable variety of network structures shown. 
Different empirical hypergraphs may exhibit quite different edge-size distributions: some, like \data{congress-bills}, have a small number of nodes but very large edges (see \autoref{tab:lp_results} below for sizes of each dataset), while others, like \data{tags-ask-ubuntu}, have much larger numbers of nodes but much smaller edges. 
The parameters from the SEM-\model fit approximate the degree and edge-size distributions for these very different empirical hypergraphs.

\subsection{Link prediction on empirical networks}

\begin{table*}
    \begin{tabular}{l l r  r r r r r r}

    Dataset & Type & $n$  & $m$  & $\bar{k}$ & AUC & F1 & AUC(t) & F1(t) \\
    \hline
    \data{ coauth-dblp } & co-author & 1,930,378 & 3,700,681 & 280 & 0.871 & 0.807 & 0.659 & 0.745 \\
    \data{ coauth-mag-geology } & co-author & 1,261,129 & 1,590,335 & 284 & 0.780 & 0.668 & 0.544 & 0.651 \\
    \data{ coauth-mag-history } & co-author & 1,034,876 & 1,812,511 & 925 & 0.639 & 0.607 & 0.367 & 0.464 \\
    \midrule
    \data{ dawn } & biological & 2,558 & 2,272,433 & 16 & $-$ & $-$ & $-$ & $-$ \\ 
    \data{ disgenenet } & biological & 12,368 & 2,261 & 2453 & 0.574 & 0.545 & NA & NA \\
    \data{ diseasome } & biological & 516 & 903 & 11 & 0.315 & 0.320 & 0.442 & 0.474\\
    \data{ kaggle-whats-cooking } & biological & 6,714 & 39,774 & 65 & 0.945 & 0.903 & NA & NA \\
    \data{ ndc-classes } & biological & 1,161 & 49,726 & 39 & 0.989 & 0.973 & 0.988 & 0.973 \\
    \data{ ndc-substances } & biological & 5,556 & 112,405 & 187 & 0.558 & 0.544 & 0.616 & 0.614 \\
    \midrule
    \data{ tags-ask-ubuntu } & webpage & 3,029 & 271,233 & 5 & 0.763 & 0.738 & 0.697 & 0.650 \\
    \data{ tags-math-sx } & webpage & 1,629 & 822,059 & 5 & 0.740 & 0.665 & 0.624 & 0.588 \\
    \data{ tags-stack-overflow } & webpage & 49,998 & 14,458,875 & 5 & $-$ & $-$ & $-$ & $-$ \\
    \data{ threads-ask-ubuntu } & webpage & 125,602 & 192,947 & 14 & 0.507 & 0.608 & 0.429 & 0.601 \\
    \data{ threads-math-sx } & webpage & 176,445 & 719,792 & 21 & 0.754 & 0.778 & 0.674 & 0.731 \\
    \data{ threads-stack-overflow } & webpage & 2,675,969 & 11,305,356 & 67 & 0.778 & 0.779 & 0.686 & 0.774  \\
    \midrule
    \data{ congress-bills } & social & 1,718 & 282,049 & 400 & 0.575 & 0.540 & 0.528 & 0.465 \\
    \data{ contact-high-school } & social & 327 & 172,035  & 5 & 0.989 & 0.947 & 0.983 & 0.937\\
    \data{ contact-primary-school } & social & 242 & 106,879 & 5 & 0.951 & 0.880 & 0.933 & 0.853 \\
    \data{ email-enron } & social & 148 & 10,885 & 37 & 0.944 & 0.882 & 0.755 & 0.707 \\
    \data{ email-eu } & social & 1,005 & 235,263 & 40 & 0.916 & 0.863 & 0.882 & 0.826 \\
    \data{ hospital-lyon } & social & 75 & 27,834 & 5 & 0.874 & 0.785 & 0.636 & 0.633 \\
    \data{ hypertext-conference } & social & 113 & 19,036 & 6 & 0.934 & 0.864 & NA & NA \\
    \data{ invs13 } & social & 92 & 9,644 & 4 & 0.968 & 0.911 & NA & NA  \\
    \data{ invs15 } & social & 232 & 73,822 & 4 & 0.928 & 0.889 & NA & NA \\
    \data{ malawi-village } & social & 86 & 99,942 & 4 & 0.99 & 0.971 & NA & NA  \\
    \data{ science-gallery } & social & 10,972 & 338,765 & 5 & 1.0 & 0.999 & NA & NA  \\
    \data{ sfhh-conference } & social & 403 & 54,305 & 9 & 0.979 & 0.922 & NA & NA \\
    \\
    \end{tabular}

    \caption{
        Link prediction on the empirical hypergraphs provided by the XGI package for Python \cite{landryXGIPythonPackage2023}. 
        The number of nodes $n$, number of edges $m$, and maximum edge size $\bar{k}$ are shown for each data set. 
        We compute area under the receiver operating characteristic (AUC) and F1 scores for models trained on timestamped sequences (t) and random sequences of edges, using $20\%$ of the total data in both cases. 
        ``NA'' indicates that timestamps were not supplied for the dataset, making it impossible to perform link prediction under the (t) condition. 
        AUC and F1 scores are computed using either the remaining $80\%$ of data or $10^5$ edges sampled uniformly at random (in the case of large datasets) as positive examples, then generating an equal number of negative examples, and forming predictions on the examples as described in the main text. 
        We did not obtain link prediction results for two of the datasets due to computational limitations in the evaluation of the marginal likelihood. 
        We set the length of $\extantnodedist$ and $\novelnodedist$ to be equal to the largest edge size for each empirical hypergraph for all runs.
        }
    \label{tab:lp_results}
\end{table*}

We now use our proposed \model to perform link prediction on the empirical hypergraphs.
At any given time, the \model assigns a probability that any given candidate edge will indeed be formed in the next timestep. 
We can therefore perform link prediction by predicting that high-probability candidates will be formed and that low-probability candidates will not.

A challenge in most link prediction contexts is the absence of true negative samples for training and evaluation. 
Because our model ``training'' is simply the SEM algorithm described above, we do not need negative samples for training; we do, however, still need them for evaluation. 
We therefore form a collection of negative examples by sampling non-realized edges in order to match the degree distribution and edge-size distribution of each empirical network. 
To do this, we draw the size for each created negative edge according to the same empirical edge-size distribution of the (positive) edges while sampling nodes to be included in the negative set based on the node degree distribution. 
To maintain balanced classes, we generate as many negative examples as there are positive examples.
In each dataset, we use $20\%$ of the observed (positive) edges to perform SEM. 
We fit our model using two distinct types of training data. 
When timestamped data is available, we take the \emph{first} $20\%$ of edges as training data. 
However, for each dataset, we also fit on a uniformly random subset of $20\%$ of the training data, which is not temporally contiguous. 
We use this strategy in order to assess the effectiveness of our model, which has explicit temporal structure, for settings in which no empirical timestamps are available.
To assess the models trained under each condition, we combine the positive and negative examples and compute the probability of each candidate being realized under the \model. 

We consider a positive prediction to be a candidate with modeled probability above the median. Because the model outputs a likelihood score (rather than a binary decision), we convert it into a binary label by using a threshold: we take the median score of the predictions over the training data as a cutoff.
This is necessary because the likelihood values produced by the model can be extremely small in practice (e.g., on the order of $10^{-2}$ or lower for edges that are highly unlikely or infeasible), and using an absolute threshold would not provide meaningful separability. 
Using this threshold, we classify candidates as positive if their modeled likelihood exceeds the median and as negative otherwise. While the binary threshold is used to compute F1, the AUC score itself is calculated by ranking candidates by their raw likelihood scores and evaluating the model's ability to separate positive from negative examples across all thresholds. 
This is possible because our model outputs a continuous score for each candidate edge.

From the model's predictions, we compute area under the receiver-operating characteristic curve (AUC) and F1 scores.
These scores are shown in \Cref{tab:lp_results}, with the (t) columns indicating the use of temporal information for model fitting. 

We use a maximum of $10^5$ edges as positive examples for evaluation; for temporally trained models we use edges immediately following the training set, while in non-temporally trained models we use up to $10^5$ edges sampled uniformly at random. 
There were two extremely dense datasets (\data{dawn} and \data{tags-stack-overflow}) in which we were able to perform inference via SEM but not able to perform link prediction due to the computational expense of computing the marginal log-likelihood of a candidate edge. 
In such hypergraphs, the number of candidate edges $f$ which could have generated a new edge $e$ is very large, leading to many terms in the marginal log-likelihood which must be computed. 

Perhaps surprisingly considering that our \model relies heavily on temporal information, the link prediction results from random draws of edges ignoring the temporal timestamp information are in many cases as good or better than the predictions obtained from using the temporal ordering supplied with the data. 
Indeed, our model was especially successful on many social interaction datasets for which no timestamps are supplied with the data (bottom rows of \Cref{tab:lp_results}). 
Despite the small number of parameters in the \model, we are able to achieve AUCs over $0.85$ in 14 out of 25 datasets for which we were able to perform link prediction.

We now compare the predictive performance of our \model to that of two neural network methods: neural network methods, Neural Hypergraph Link Prediction (NHP) \cite{yadatiNHPNeuralHypergraph2020} and the Logical Hyperlink Predictor (LHP) \cite{yangLHPLogicalHypergraph2023}.
The NHP study \cite{yadatiNHPNeuralHypergraph2020} shows results on 5 datasets and the LHP study \cite{yangLHPLogicalHypergraph2023} shows results on 4 datasets. 
We test our \model on the three of these datasets that appear in both papers, replicating the experimental setup from these studies as closely as possible. 
We compute AUC and F1 scores for link prediction, as well as training time, and compare our results to the published results for NHP and LHP.
To ensure comparability, we follow the same training setup: we use a uniformly random $20\%$ of hyperedges for training and use the rest for testing. 
Here, in order to ensure a fair comparison with LHP and NHP, we do not use the previously described degree and size matched sampling method, but instead adopt their approach to negative sampling: for each positive hyperedge, we create a corresponding negative sample by randomly replacing half of the nodes in the edge with nodes from the rest of the hypergraph.
Because these datasets are relatively small, we set both $\extantnodedist$ and $\novelnodedist$ to a maximum length of $\bar{k} = 5$ during training. 
Combined with $\eta$, this gives our model a total of 11 scalar parameters. 
The neural network models are much larger: while we do not have exact published parameter counts, information in the LHP paper gives at least $9.8 \times 10^5$ scalar parameters.

We observe in \Cref{tab:benchmarking} that our \model performs competitively with the two neural network models: on the iJO1366 dataset it substantially outperforms both; on the iAF1260b dataset it performs better than NHP and worse than LHP; and on the USPTO dataset it performs worse than both NHP and LHP. 
We conjecture that the comparatively poor performance of our model reflects in part the small maximum edge size of this data set compared to the other two. Another important consideration is the parameter count and training efficiency of neural network models. Specifically, for larger networks such as \data{threads-stack-overflow}, neural network methods become infeasible due to the high number of nodes and edges, whereas \model remains scalable and efficient.
Although our training times are typically larger than those reported for LHP, we note that LHP was trained on a TITAN RTX GPU while our model was trained on X. H.'s personal laptop, which possesses an 11th Gen Intel i7-11800H processor with 8 cores and 16 logical processors and 16GB of RAM.



\begin{table*}

    \resizebox{\textwidth}{!}{%
        \begin{tabular}{l | r r r | r r r | r r r | r r r}
 & $n$ & $m$ & Max Edge Size & NHP AUC & NHP F1 & NHP T & LHP AUC & LHP F1 & LHP T & \model AUC & \model F1 & \model T \\
\hline
iAF1260b & 1,668 & 2,084 & 67 & 0.582 & 0.415 & 1m22s & 0.639 & 0.588 & 0m12s & 0.605 & 0.539 & 0m42s  \\
iJO1366 & 1,805 & 2,253  & 106 & 0.599 & 0.400 & 1m50s & 0.638 & 0.620 & 0m17s & 0.774 & 0.786 & 0m48s \\
USPTO & 16,293 & 11,433 & 8 & 0.662 & 0.500 & 6m15s & 0.733 & 0.650 & 1m35s & 0.515 & 0.554 & 0m16s \\
\end{tabular}
        }

    \caption{
        Link prediction results of our \model against two neural network hypergraph link prediction methods, NHP \cite{yadatiNHPNeuralHypergraph2020} and LHP \cite{yangLHPLogicalHypergraph2023}. 
        The iAF1260b and iJO1366 datasets are metabolic reaction hypergraphs while USPTO is an organic reactions hypergraphdataset. 
        Negative edges are generated according to the description and code for LHP. 
        Since none of these datasets are temporal hypergraphs, results are calculated from $20\%/80\%$ training-testing splits over 10 independent randomized trials. 
        We set the lengths of both $\extantnodedist$ and $\novelnodedist$ to be $5$ for all runs. 
        The reported AUC and F1 scores for LHP and NHP are taken from Table 4 of the LHP paper \cite{yangLHPLogicalHypergraph2023}. 
        The original NHP paper presents lower AUC scores compared to those reported in the LHP paper, possibly due to randomization; the higher of the two reported scores were used here. 
        The times to train the models (``T'') are also taken from the LHP paper (Table 7), which used an NVIDIA TITAN RTX GPU. 
        Time for fitting the \model parameters was measured on X. H.'s personal computer, which possesses a a 11th Gen Intel i7-11800H that has 8 cores and 16 logical processors and 16GB of RAM. 
}
\label{tab:benchmarking}
\end{table*}

\section{Discussion} \label{sec:discussion}

    We proposed the \Model, a simple model of hypergraph evolution based on a noisy edge-copying mechanism. 
    This model is mechanistic, interpretable, and analytically tractable. 
    In addition to several analytic descriptions of the model's behavior, we also provide a scalable stochastic expectation maximization algorithm for fitting the model to empirical data.
    We find in \Cref{tab:benchmarking} that our 11-parameter model is competitive on link prediction tasks with neural network models containing hundreds of thousands of parameters.

    While we primarily evaluate link prediction using global metrics such as AUC, recent work \cite{menand2024link} highlights that such metrics can overestimate performance in certain settings. Vertex-centric evaluation frameworks may reveal additional insights into the local predictive performance of algorithms. Greater attention should be given to vertex-centric metrics, which represent a valuable complement to global evaluation methods.
    
    One direction of future work concerns modeling of recombination of edges. 
    In our \model, new edges are formed from a noisy copy of a single prior edge. 
    It may be of interest to allow edges to to form from multiple prior edges; such a process might model, for example, the formation of broad collaborations from multiple research groups. 
    A version of this idea is explored in the context of sequences of sets by   Benson et al. \cite{bensonSequencesSets2018}. 
    Incorporating such a mechanism into the \model would significantly complicate the inference problem, since the set of possible generators of each edge would grow large. 
    Fitting such a model to data might require more sophisticated computational techniques that could be of independent interest. 
    Additionally, custom hypergraph data structures tailored to specific applications can significantly boost computational efficiency and can be a promising future direction.
    
    Another direction of future work involves the incorporation of hypergraph metadata. 
    In a hypergraph with node attributes, for example, one might model an edge sampling step in which the node $v_0$ sampled from the selected edge $e$ acts as a ``leader'' in the next edge; other edges in $e$ could be more likely to be copied into the next edge if they share attributes with $v_0$.
    
    It is often claimed that the widespread availability of large data sets and computational power sufficient to train deep models has made theory and simple models obsolete in predictive tasks \cite{anderson2008end}.
    We take our results to suggest a continuing role for simple, interpretable stochastic models in the study of complex systems, even in the age of deep learning.

\subsection{Software}
We implemented our model and performed experiments using the XGI package \cite{landryXGIPythonPackage2023} for higher-order network analysis in  Python. 
The XGI package also makes available the datasets used in our experiments. (We conducted experiments using all datasets provided in the XGI package.)
Code sufficient to fully reproduce our experiments is available at \url{https://github.com/hexie1995/HyperGraph/tree/prod}.

\subsection{Acknowedgements}
We thank Rebecca Hardenbrook, Nicholas Landry, Violet Ross, Alice Schwarze, and Anna Vasenina for helpful discussions. X.H. and P.J.M. were supported by the Army Research Office under MURI award W911NF-18-1-0244 and the National Institutes of Health FIC under R01-TW011493. P.S.C. was supported by the National Science Foundation under DMS-2407058. P.J.M. was additionally supported by the National Science Foundation under BCS-2140024 and DEB-2308460.

\bibliographystyle{plain}
\bibliography{pc-automated-refs.bib}

\newpage

\appendix

\section{Parameter estimation in synthetic HCM hypergraphs.}
\begin{figure*}[h!]
    \centering
    \includegraphics[scale=0.55]{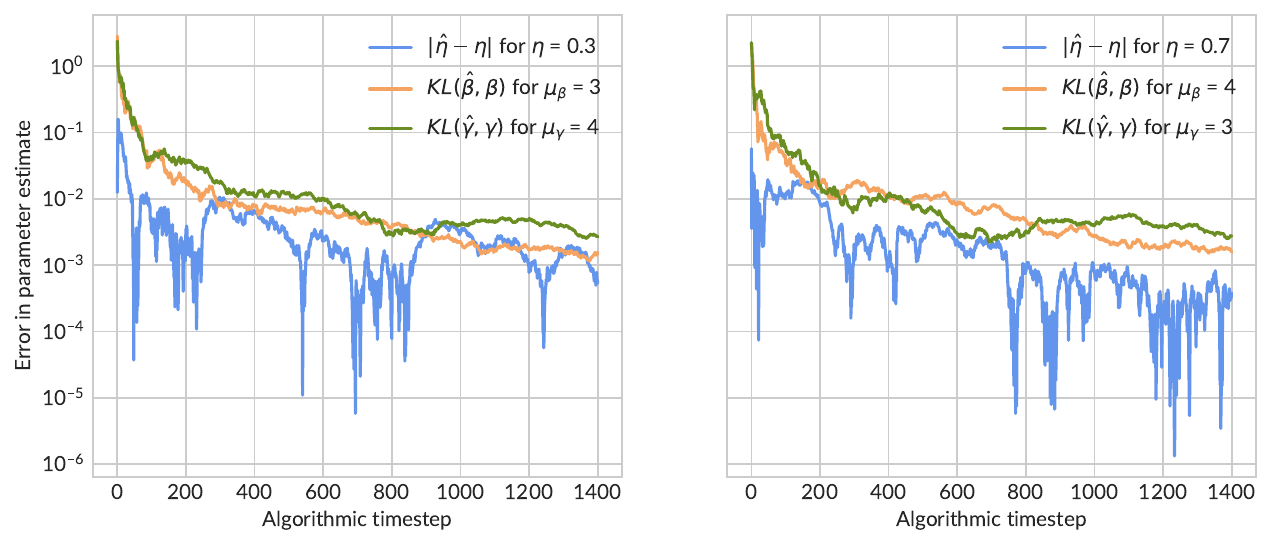}
    \caption{Estimation of parameters in two synthetic hypergraphs with 10k hyperedges 
    each. The convergence criterion, tested every 100 steps, was reached at 1400 steps for each of them. 
    We show the absolute difference $\abs{\hat{\eta} - \eta}$ between the estimate and true value of $\eta$. 
    We also show the Kullback--Leibler divergence of the estimates of $\vbeta$ and $\vgamma$ from the true values.    
    The $\beta$ and $\gamma$ here are generated with a random Poisson distribution, truncated at value $10$, with means $\mu_{\vbeta}$ and $\mu_{\vgamma}$ indicated in the plot legends. 
    Note the logarithmic scale of the vertical axis.}
    \label{fig:accuracy}
\end{figure*}

For consistency, we applied the same convergence criterion across all experiments in the paper. Specifically, every 100 steps, we calculated the difference between the current and previous $\eta$ values. If the percentage change was less than $1\%$, the loop was terminated. As examples, in \Cref{fig:accuracy} we show the errors in the parameter estimates as obtained after different numbers of steps of the algorithm for two synthetic hypergraphs.

\begin{table*}[h!]
    \centering
     \begin{tabular}{ l r r r r r }
    &        &               &                & Clock         & Training \\
    &        &               &                & time    & steps \\
    & $\eta$ & $\mu_{\beta}$ & $\mu_{\gamma}$ & (s) & ($\times 10^3$)\\
\hline
\data{coauth-dblp} & 0.505 & 0.403 & 0.512 & 23.6 & 11 \\
\data{coauth-mag-} & 0.365 & 0.585 & 0.630 & 8.3 & 8 \\
\data{coauth-mag-history} & 0.242 & 0.202 & 0.110 & 14.0 & 18 \\
\midrule
\data{dawn} & 0.686 & 4.15E-19 & 0.071 & 8064.3 & 4 \\
\data{disgenenet} & 0.099 & 6.90E-18 & 6.832 & 30.3 & 5 \\
\data{diseasome} & 0.078 & 6.23E-04 & 0.537 & 1.6 & 10 \\
\data{kaggle-whats-cooking} & 0.393 & 0.151 & 5.019 & 930.7 & 5 \\
\data{ndc-classes} & 0.996 & 0.009 & 0.019 & 41.1 & 3 \\
\data{ndc-substances} & 0.680 & 0.014 & 0.180 & 87.0 & 5 \\
\midrule
\data{tags-ask-ubuntu} & 0.614 & 0.005 & 0.569 & 434.7 & 4 \\
\data{tags-math-sx} & 0.785 & 3.02E-04 & 0.197 & 1927.5 & 5 \\
\data{tags-stack-overflow} & 0.773 & 0.003 & 0.372 & 29212.2 & 4 \\
\data{threads-ask-ubuntu} & 0.058 & 0.477 & 0.468 & 46.6 & 13 \\
\data{threads-math-sx} & 0.165 & 0.191 & 0.848 & 85.8 & 4 \\
\data{threads-stack-overflow} & 0.082 & 0.149 & 1.007 & 92.6 & 6 \\
\midrule
\data{congress-bills} & 0.294 & 0.003 & 1.379 & 563.2 & 5 \\
\data{contact-high-school} & 0.975 & 1.16E-04 & 0.034 & 81.0 & 3 \\
\data{contact-primary-school} & 0.910 & 8.25E-05 & 0.098 & 111.5 & 5 \\
\data{email-enron} & 0.887 & 0.013 & 0.083 & 24.8 & 3 \\
\data{email-eu} & 0.910 & 0.002 & 0.108 & 205.2 & 5 \\
\data{hospital-lyon} & 0.945 & 3.86E-05 & 0.069 & 148.9 & 4 \\
\data{hypertext-conference} & 0.888 & 2.63E-18 & 0.109 & 34.1 & 3 \\
\data{invs13} & 0.917 & 9.77E-08 & 0.082 & 23.8 & 4 \\
\data{invs15} & 0.938 & 4.11E-06 & 0.065 & 56.8 & 3 \\
\data{malawi-village} & 0.996 & 1.32E-18 & 0.005 & 168.8 & 3 \\
\data{science-gallery} & 0.871 & 2.10E-04 & 0.121 & 6.7 & 4 \\
\data{sfhh-conference} & 0.827 & 1.97E-20 & 0.185 & 69.3 & 5 \\
\end{tabular}

    \caption{Details of parameters retrieved through stochastic expectation maximization with batch size 30 for each hypergraph. We set both the length of $\beta$ and $\gamma$ to be equivalent to the largest edge size in the corresponding real-world hypergraph.}
    \label{tab:parameter-values}
\end{table*}

\section{Reconstruction of Hypergraph Properties from Generative Models}

\begin{figure*}[h!]
    \centering
    \includegraphics[scale=0.55]{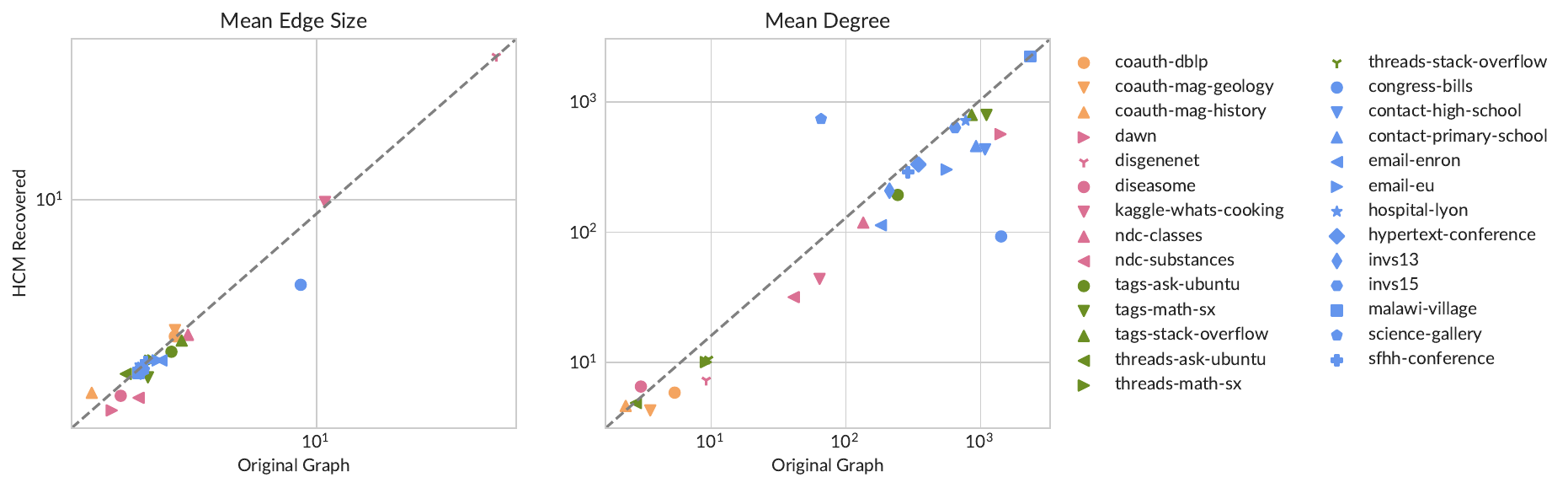}
    \caption{Mean edge sizes and mean node degrees compared between fit \model instances and empirical data.
    Note the log-log scale.}
    \label{fig:Mean_sizes}
\end{figure*}

In \autoref{fig:Mean_sizes}, we compare the mean edge size and mean degree between hypergraphs generated from estimated parameters (listed in \autoref{tab:parameter-values}) and the original data, for a variety of real-world hypergraphs.

\begin{figure*}[!ht]
    \centering
    \includegraphics[width=1\linewidth]{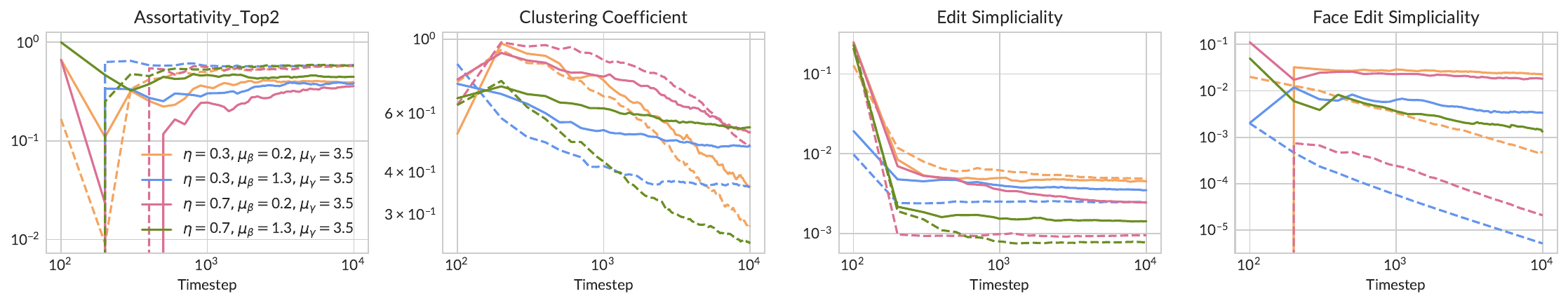} 
    \caption{Hypergraph properties including top-2 degree assortativity, clustering coefficient, edit simpliciality, and face edit simpliciality from different generative models. Solid lines show values for data generated by the \model. Dashed lines are for Erd\H{o}s-R\'enyi hypergraphs generated with the same expected edge-size. The horizontal axis ranges from $t=0$ to $10$k, showing how the properties change with the numbers of steps in the growing hypergraphs.}
    \label{fig:four-properties-supp}
\end{figure*}

We perform a similar analysis for synthetic datasets, comparing different hypergraph properties, including face edit simplicity, which is a more localized concept indicating the number of subfaces needed to be added to the hypergraph to render a specific face a simplex.  
We observe minimal differences in degree assortativity and edit simplicity between the \model and ER hypergraphs in \autoref{fig:four-properties-supp}. However, discrepancies arise in the clustering coefficient and face edit simplicity. Particularly, we see the face edit simpliciality for ER hypergraphs dropped persistently over time, which is very different behavior from the \model (except in the case when $\eta=0.7$ and $\mu_{\novelnodedist}=1.3$, and even in that case the \model face edit simpliciality is orders of magnitude larger than for the corresponding ER hypergraphs). As was shown in \cite{landrySimplicialityHigherorderNetworks2024}, face edit simpliciality is usually higher than $10^{-2}$ for real-world hypergraphs, and thus the real-world replication of the ER model is much worse than that of our \model, which keeps steadily above $10^{-2}$ for all cases studied. 

For the clustering coefficient, interestingly, we see in \autoref{fig:four-properties-supp} that the values from the ER and \model are almost the same when $\mu_{\novelnodedist}$ is small. However, when $\mu_{\novelnodedist} = 1.3$, the clustering coefficient quickly drops for the ER hypergraphs, but stays steady for the HCM model, which has values close to the real world hypergraph clustering coefficients described in \cite{gallagherClusteringCoefficientsProtein2013}. These properties of our \model with various parameter sets demonstrate the ability of the model to generate realistic-seeming hypergraphs, despite its small number of parameters. 

\begin{figure*}[!ht]

\subfloat[\label{fig:additionalM4a}]{%
    \includegraphics[width=\textwidth]{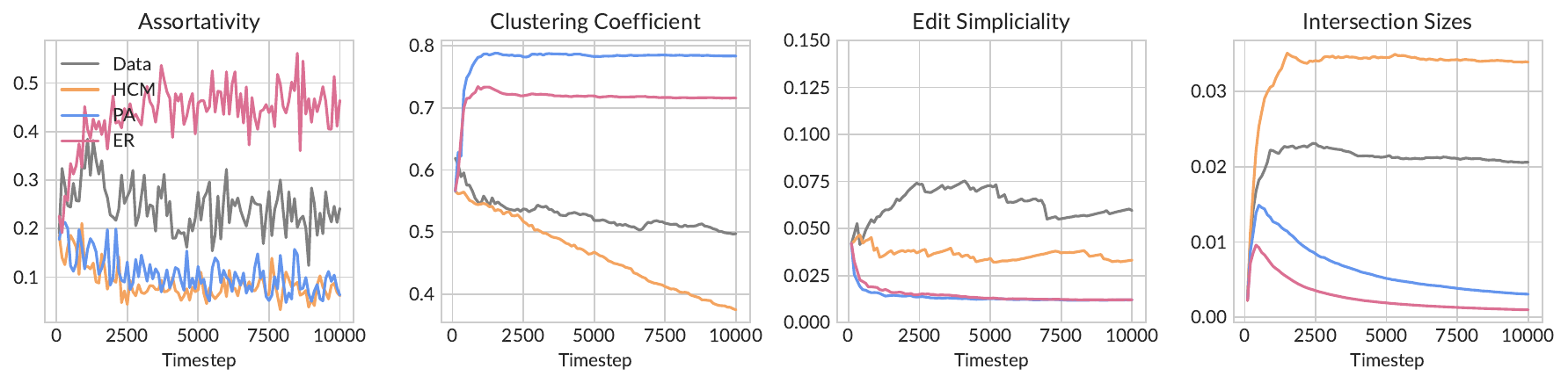}
}

\subfloat[\label{fig:additionalM4b}]{%
    \includegraphics[width=\textwidth]{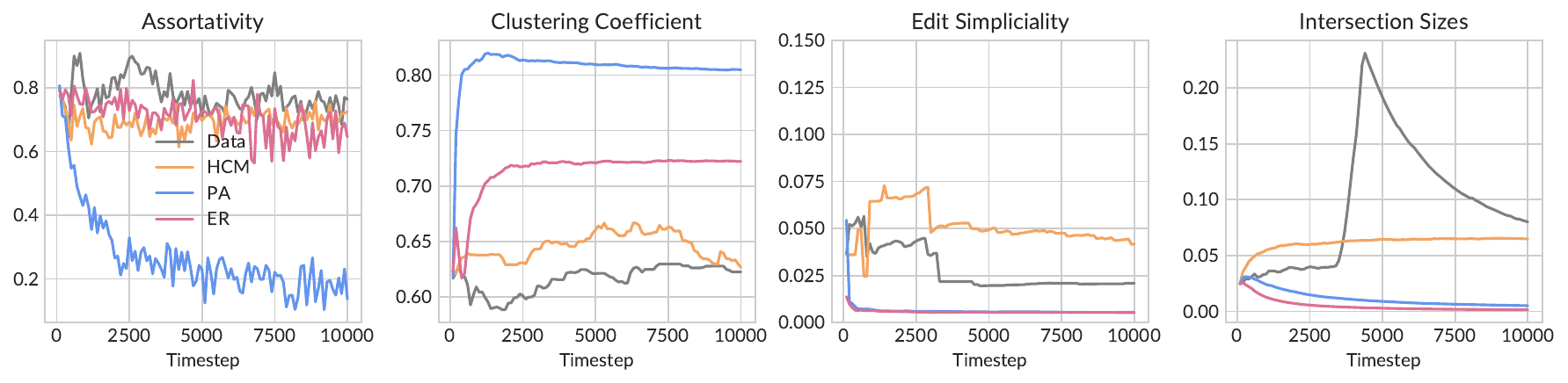}
}

\subfloat[\label{fig:additionalM4c}]{%
   \includegraphics[width=\textwidth]{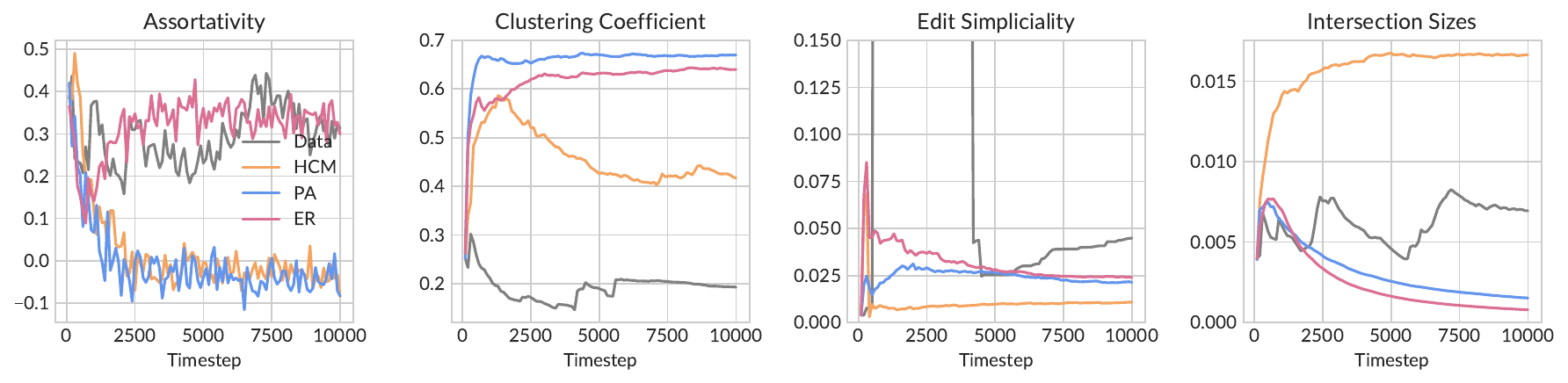}
}

\caption{Structural properties of the \data{email-eu} (a), \data{ndc-classes} (b), and \data{ndc-substances} (c) data sets. We show that these datasets have exhibited behaviors distinct from those of \data{email-enron}, shown in \Cref{fig:four-properties} in the main text, highlighting that assortativity is highly dependent on the underlying structure of the real-world hypergraph. We also note that HCM is the only model that demonstrates non-diminishing edge intersection sizes across the different datasets.}
    \label{fig:additionalM4}
\end{figure*}

In \Cref{fig:additionalM4}, we repeat the experiment shown in  \Cref{fig:four-properties} for \data{email-enron} on other datasets, to demonstrate different exhibited behaviors across these other datasets, computing the assortativity, clustering coefficient, edit simpliciality, and intersection sizes for each dataset and the corresponding \model, ER and PA models.

\section{Asymptotic Properties}

\subsection{Power-law degree distribution} \label{sec:power-law-tails}

We now give a heuristic argument that the degree distribution of our model has an asymptotic power-law tail, with an exponent that depends on the model parameters. 
Our argument is based on rate equations, generalizing an argument by Mitzenmacher \cite{mitzenmacherBriefHistoryGenerative2004} in the context of preferential attachment dyadic graphs. 

Fix $d > 1$. 
Let $\at{p}{t}_d$ be the proportion of nodes which have degree $d$ at time $t$. 
The total number of such nodes is $\at{n}{t} \at{p}{t}_d$. 
In timestep $t+1$, the change in the number of these nodes is $\at{\Delta}{t+1}_d = \at{n}{t+1} \at{p}{t+1}_d - \at{n}{t} \at{p}{t}_d$. 
We now estimate $\at{\delta}{t+1}_d = \mathbb{E}\brackets{\at{\Delta}{t+1}_d}$ in expectation. 
The probability of an individual node being selected in the edge-sampling step is proportional to its degree $d$ in the current hypergraph $\at{\hypergraph}{t}$, since there are $d$ edges which could be selected which contain that node. 
Normalizing, the probability that a given node selected in the edge-sampling step has degree $d$ is $\frac{d}{\mean{\at{d}{t}}}\at{p}{t}_d$, where $\mean{\at{d}{t}}$ is the mean degree at time $t$. 
When a node of degree $d$ is selected in the edge-sampling step it becomes a node of degree $d+1$, while when a node of degree $d-1$ is selected it becomes a node of degree $d$. 
The expected number of nodes so selected is $1 + \eta\paren{\mean{\at{k}{t}}-1}$, where $\mean{\at{k}{t}}$ is the mean edge size at time $t$. 
Thus, the expected number of degree $d$ nodes created through the edge-sampling step is $\frac{1 + \eta\paren{\mean{\at{k}{t}}-1}}{\mean{\at{d}{t}}}\brackets{\paren{d-1}\at{p}{t}_{d-1} - d\at{p}{t}_d}$. 
Through similar reasoning, the expected number of degree $d$ nodes created through the process of extant node addition is $\mu_{\extantnodedist}\brackets{\at{p}{t}_{d-1} - \at{p}{t}_d}$.
Since we have assumed $d > 1$ but novel node addition can only create nodes of degree $1$, only these two processes contribute. 
Our expected rate equation is then 

\begin{widetext}
\begin{align}
    \at{n}{t+1} \at{p}{t+1}_d - \at{n}{t} \at{p}{t}_d &=  \frac{1 + \eta\paren{\mean{\at{k}{t}}-1}}{\mean{\at{d}{t}}}\brackets{\paren{d-1}\at{p}{t}_{d-1} - d\at{p}{t}_d}  + \mu_{\extantnodedist}\brackets{\at{p}{t}_{d-1} - \at{p}{t}_d}\;. \label{eq:rate-equation}
\end{align}
\end{widetext}

We now assume stationarity of the degree distribution and the edge size sequence. 
At stationarity, we must have $\at{p}{t+1}_d = \at{p}{t}_d = p_d$ for some constant $p_d$, as well as $\mean{\at{k}{t}} = \mean{k}$ and $\mean{\at{d}{t}} = \mean{d}$ for constants $\mean{k}$ and $\mean{d}$. 
We also have that $\mean{d} = \frac{m}{n}\mean{k} = \frac{\mean{k}}{\mu_{\novelnodedist}}$, where we have used the fact that $\mu_{\novelnodedist}$ nodes are added per edge in each timestep. 
Finally, in expectation, $\at{n}{t+1} - \at{n}{t} = \mu_{\novelnodedist}$.
At stationarity and in expectation, our approximate compartmental equation now reads 
\begin{align}
    \mu_{\novelnodedist} p_d = a\brackets{\paren{d-1}p_{d-1} - dp_d} + \mu_{\extantnodedist}\brackets{p_{d-1} - p_d}\;, \label{eq:stationary-rate-equation}
\end{align}
where we have defined $a = \frac{1 + \eta\paren{\mean{k}-1}}{\mean{d}}$. 
Rearrangement yields 
\begin{align}
    \frac{p_d}{p_{d-1}} &= \frac{a(d-1) + \mu_{\extantnodedist}}{ad + \mu_{\extantnodedist} + \mu_{\novelnodedist}} \\ 
    &= 1 - \frac{a + \mu_{\novelnodedist}}{ad + \mu_{\extantnodedist} + \mu_{\novelnodedist}}\;.
\end{align}
Since we are interested in the tails of the degree distribution, we allow $d$ to grow large, yielding 
\begin{align}
    \frac{p_d}{p_{d-1}} &\approx 1 - \frac{a + \mu_{\novelnodedist}}{a}\frac{1}{d} \\ 
    &\approx \paren{\frac{d-1}{d}}^{1 + \frac{\mu_{\novelnodedist}}{a}}\;. 
\end{align}
Unfolding the recurrence yields, for large $d$, a power-law tail, $p_d\sim d^{-\zeta}$, with exponent $\zeta = 1 + \frac{\mu_{\novelnodedist}}{a}$.
Inserting explicit formulae for $a$, $\mean{k}$ (\cref{eq:mean-edge-size}) and $\mean{d}$ (\cref{eq:mean-degree}), 
we can express this exponent explicitly in terms of the model parameters using the formulas for $a$, $\mean{k}$, and $\mean{d}$, obtaining 
\begin{align}
    \zeta = 1 + \frac{1 - \eta + \mu_{\extantnodedist} + \mu_{\novelnodedist}}{1 - \eta(1 - \mu_{\extantnodedist} - \mu_{\novelnodedist})}\;. \label{eq:power-law-exponent}
\end{align}

\subsection{Edge-size distribution} \label{sec:edge-sizes}

We describe the entries of the matrix $\mathbf{W}$ described in the main text. 
Each entry $w_{ij}$ of $\mathbf{W}$ represents the conditional probability 
\begin{align}
    w_{ij} = \prob{\abs{e} = i \given \abs{f} = j}\;,
\end{align}
where $f$ is the edge sampled in the edge-selection step of \Cref{alg:model} and $e$ is the final edge formed.  
We explicitly construct these probabilities conditional on the events that $\ell+1$ nodes are selected in the edge-sampling step and that $h$ nodes are selected in the extant node addition step. 
With these conditional probabilities, the law of total probability then yields 

\begin{widetext}
\begin{align}
    w_{ij} = \prob{\abs{e} = i \given \abs{f} = j} 
           = \sum_{\ell = 0}^{j-1} \sum_{h = 0}^{i - \ell} \binom{j-1}{\ell} \eta^{\ell} (1-\eta)^{j - \ell - 1} \extantnodepar_h \novelnodepar_{i - \ell - h}\;.
\end{align}
\end{widetext}

The Perron eigenvector of this matrix gives the stationary distribution of edge sizes in the model. 
In practice, it is necessary to select a finite size for the matrix $\mathbf{W}$, which amounts to artificially setting $w_{ij} = 0$ for $i$ and $j$ sufficiently large.


\subsection{Asymptotic properties of intersection sizes} \label{sec:intersection-sizes}

We now develop a probabilistic argument supporting the following claim: 
\begin{clm*}
    Let $r_{ijk}$ be the expected proportion of pairs of the $m$ edges, relative to the total number of pairs $\binom{m}{2}$, which have sizes $i$ and $j$ with intersection size $k$,  with the expectation taken with respect to our proposed model and with the edge of size $j$ appearing before the edge of size $i$. 
    Then, there exist constants $q_{ijk} \geq 0$ such that: 
    \begin{align}
        r_{ijk} = \begin{cases}
            q_{ijk} + o(1) &\quad k = 0 \\ 
            m^{-1}q_{ijk} + O(m^{-2}) &\quad k \geq 1\;. 
        \end{cases}
    \end{align}
    Furthermore, the scalars $q_{ijk}$ can be approximated as the solution of an eigenvector problem $\mathbf{q} = \mathbf{C}\mathbf{q}$, where $\mathbf{q}$ collects the scalars $q_{ijk}$ in flattened form and the matrix $\mathbf{C}$ is a function of the model parameters $\alpha$, $\vbeta$, and $\vgamma$.
\end{clm*}

Throughout this section, we fix the timestep $t$. 
We use the shorthand $z \eqdef \at{z}{t}$ and $z' \eqdef \at{z}{t+1}$ for any time-dependent quantities $z$. 
We also say that $f(m) \doteq g(m)$ if $\lim_{m\rightarrow \infty} \frac{f(m)}{g(m)} = 1$, where $m$ is the number of edges in $\cH$. 
We write $g \prec e$ if $e$ was added to $\cH$ after $g$.

Let $R_{ijk} \eqdef \prob[t]{\abs{e} = i, \abs{g} = j, \abs{e\cap g} = k \given g \prec e}$, where 
the probability is computed with respect to the empirical distribution of $\at{\cE}{t}$. 
The quantity $R_{ijk}$ corresponds to a sampling process in which we uniformly select a pair of edges; arrange them in descending order according to the timestep in which they were sampled; label the first (later) edge $e$ and the second (earlier) edge $g$; and then check whether $\abs{e} = i$, $\abs{g} = j$, and $\abs{e\cap g} = k$.
We will study the limiting behavior of $r_{ijk} \eqdef \E{R_{ijk}} $ as $t \rightarrow \infty$. 
Let $m = \at{m}{t}$ and assume that one edge is added every timestep. 
Let $P_{ijk} \eqdef \binom{m}{2} R_{ijk}$ be the corresponding number of ordered pairs of edges of sizes $\abs{e} = i$ and $\abs{g} = j$ with intersection size $k$, ordered so that $g \prec e$.
Let $p_{ijk} = \E{P_{ijk}}$.

We first write down a compartmental update for $P_{ijk}$ when a single edge $e$ is added. 
Assume that $e$ is sampled from edge $f$ in the edge-sampling step. 
Let $Z_{eg, ijk}$ be the indicator of the event that the new edge $e$ has size $i$, a previously existing edge $g$ has size $j$, and the intersection size of $e$ and $g$ has size $k$. 
Then, the compartmental update for $P_{ijk}$ reads
\begin{align}
    P'_{ijk} = P_{ijk} + \sum_{g \in \cE} Z_{eg, ijk}\;. 
\end{align}
It is useful to condition on whether $g = f$, the edge that was sampled in generating $e$: 
\begin{align}
    P'_{ijk} = P_{ijk} + Z_{ef, ijk} + \sum_{g \in \cE\setminus f} Z_{eg, ijk}\;. 
\end{align}
Computing expectations gives 
\begin{align}
    p'_{ijk} = p_{ijk} + z_{ef, ijk} + \sum_{g \in \cE\setminus f} z_{eg, ijk}\;, 
\end{align}
where we have defined $z_{eg, ijk} = \E{Z_{eg, ijk}}$.
Since $z_{eg, ijk}$ only depends on edge $g$ through its size $j$ whenever $g\neq f$, let us write $y_{ijk} \eqdef z_{eg, ijk}$. 
Similarly, we will use the simplifying notation $z_{ijk} \eqdef z_{ef,ijk}$. 
Our expected compartmental update becomes 
\begin{align}
    p'_{ijk} = p_{ijk} + z_{ijk} + (m-1) y_{ijk}\;. \label{eq:compartmental-1}
\end{align}
We aim to close \cref{eq:compartmental-1} approximately by expressing $z_{ijk}$ and $y_{ijk}$ in terms of $r_{ijk}$. 

Let us first consider $z_{ijk}$. 
The probability that an edge $f$ of size $j$ is selected uniformly at random for sampling can be written 
in terms of $r_{ijk}$ by total probability, summing appropriately over the possible sizes of some other edge $\tilde{f}\neq f$ and conditioning on whether $\tilde{f}$ appears before or after $f$, noting that $\prob{f \prec \tilde{f}} = \prob{\tilde{f} \prec f} = \frac{1}{2}$ in the absence of any information about the time step corresponding to edge $f$, giving
\begin{widetext}
\begin{align}
    r_j &\eqdef \prob{\abs{f}=j} \\
    &= \sum_{\ell h} \prob{\abs{\tilde{f}} = \ell, \abs{f} = j, \abs{f\cap \tilde{f}} = h \given f \prec \tilde{f}} \prob{f \prec \tilde{f}} \\
    &\quad+ \sum_{\ell h}\prob{\abs{f} = j, \abs{\tilde{f}} = \ell, \abs{f\cap \tilde{f}} = h \given \tilde{f} \prec f} \prob{\tilde{f} \prec f} \\
    &= \frac{1}{2}\sum_{\ell h} \paren{r_{\ell jh } + r_{j \ell h }}\;.
\end{align}
\end{widetext}
Given $\abs{f}=j$, the probability that the newly-formed edge $e$ has size $i$ and that its intersection with $f$ has size $k$ is then 
\begin{widetext}
\begin{align}
    b_{ik|j} &\eqdef \prob{\abs{e} = i, \abs{e \cap f} = k \given \abs{f} = j}  \\ 
    &= \prob{\abs{e \cap f} = k \given \abs{f} = j} \prob{\abs{e} = i \given  \abs{e \cap f} = k,  \abs{f} = j} \\ 
    &= \prob{\abs{e \cap f} = k \given \abs{f} = j} \prob{\abs{e} = i \given  \abs{e \cap f} = k}\\ 
    &= \binom{j-1}{k-1}\eta^{k-1}(1-\eta)^{j-k} \sum_{x = 0}^{i-k} \beta_x \gamma_{i - k - x}\;, \label{eq:bikj}
\end{align}
\end{widetext}
where the third line follows from the second because, for large graphs, the role of the sampled edge $f$ in determining the properties of the new edge $e$ is fully captured by their intersection.
Importantly, this expression does not depend on $m$ in the long-time limit, since $m$ is large enough so that the number of nodes $n$ in the hypergraph is at least $j - k$. We also note that we assume $\beta_0<1$. 
We therefore have 
\begin{align}
    z_{ijk} &= b_{ik|j} r_j
            = \frac{1}{2} b_{ik|j} \sum_{\ell h} \paren{r_{\ell jh } + r_{j \ell h }}\;.  \label{eq:zijk}
\end{align}

We now study $y_{ijk}$. 
Let us condition on $\abs{f} = \ell$, $\abs{f \cap g} = h$, and the relative temporal order $f \prec g$ v.\ $g \prec f$.
Noting again that $\prob{f \prec g} = \prob{g \prec f} = \frac{1}{2}$, we have

\begin{widetext}
\begin{align}
    y_{ijk} &\eqdef \prob{\abs{e} = i, \abs{g} = j, \abs{e \cap g} = k}  \\
            &= \sum_{\ell h} \prob{g \prec f} \underbrace{\prob{\abs{f} = \ell, \abs{g} = j, \abs{f \cap g} = h\given g \prec f}}_{=r_{\ell j h}} \underbrace{\prob{\abs{e} = i,  \abs{e \cap g} = k \given\abs{f} = \ell, \abs{g} = j,  \abs{f \cap g} = h, g \prec f}}_{\eqdef a^\succ_{ik|\ell j h}}  \\ 
            &\quad+ \sum_{\ell h} \prob{f \prec g} \underbrace{\prob{\abs{g} = j, \abs{f} = \ell,  \abs{f \cap g} = h\given f \prec g}}_{=r_{ j \ell h}} \underbrace{\prob{\abs{e} = i,  \abs{e \cap g} = k \given \abs{g} = j,  \abs{f} = \ell,  \abs{f \cap g} = h, f \prec g}}_{\eqdef a^\prec_{ik|j\ell h}} \\ 
            &= \frac{1}{2}\sum_{\ell h} \paren{r_{\ell j h} a^\succ_{ik|\ell jh} + r_{j \ell h} a^\prec_{ik|j\ell h}}\;,
\end{align}
\end{widetext}
where the superscript $\{\succ,\prec\}$ on the $a$ coefficients indicates whether $e$ originates as a copy of the succeeding or preceding edge in the pair, respectively. In our calculation of these $a$ coefficients below, they will be equivalent at the level of the present approximation. However, it will be important to note that, unlike $b_{ik|j}$, these $a$ coefficients depend on $m$. 
For notational compactness, let $I^\succ_{\ell jh}$ be the event $\braces{\abs{f} = \ell, \abs{g} = j,  \abs{f \cap g} = h, g \prec f}$ and $I^\prec_{j\ell h}$ be the event $\braces{\abs{g} = j,  \abs{f} = \ell,  \abs{f \cap g} = h, f \prec g}$, again denoting whether the distinguished edge $f$ to be copied is the succeeding or preceding edge in the pair (and the index on the event $I$ continuing our convention of listing first the size of the succeeding edge, then the size of the preceding edge, and lastly the size of their intersection). 

Under this notation, we have
\begin{equation}
    a^*_{ik|\ell j h} \eqdef \prob{\abs{e} = i, \abs{e \cap g} = k \given I^*_{\ell j h}}
\end{equation}
where the $*$ is either $\succ$ or $\prec$ to distinguish the two cases.
Let $S(e)$ be the number of nodes in $e$ formed by the edge-sampling step.
Then, $\abs{e \cap f} = S(e)$. 
We expand $a^*_{ik|\ell j h}$ by conditioning on $s = S(e)$, writing 
\begin{widetext}
\begin{align}
    a^*_{ik|\ell j h} &\eqdef \prob{\abs{e} = i, \abs{e\cap g} = k \given I^*_{\ell j h}} 
    = \sum_{s} \prob{S(e) = s \given I^*_{\ell j h}} \prob{\abs{e} = i, \abs{e\cap g} = k \given S(e) = s , I^*_{\ell j h}}  \nonumber\\ 
    &= \sum_{s} \prob{S(e) = s \given I^*_{\ell j h}} \prob{\abs{e\cap g} = k \given S(e) = s , I^*_{\ell j h}} \prob{\abs{e} = i \given  S(e) = s ,  \abs{e\cap g} = k , I^*_{\ell j h}}  \;, \label{eq:a-star-1}
\end{align}
\end{widetext}
where we take note to observe that the sum over possible values of $s=S(e)$ here ranges from $0$ to $\min(|e|,|g|)$, where $|g|=j$ when calculating $a^\succ_{ik|\ell j h}$ and $|g|=\ell$ when calculating $a^\prec_{ik|\ell j h}$.

Introducing additional notation,  
let $S(e,g)$ be the number of nodes in $e$ \emph{formed by the edge-sampling step} that are also elements of edge $g$. 
Under our model definition, this is equivalent to  $S(e,g)= \abs{e \cap f \cap g}$. 
Similarly, let $X(e, g)$ be the number of nodes in $e$ \emph{formed by the extant node addition step} that are also elements of edge $g$.
Abusing notation, we also let $X(e)$ denote the total number of nodes in $e$ formed through the extant node addition step, regardless of whether they intersect with any other edges. 
Then, for any edge $g \prec e$, we have $\abs{e \cap g} = S(e,g) + X(e,g)$.
We note that under the HCM step process, $S(e,g)$ is independent of $X(e,g)$ and $S(e)$ is independent of $X(e)$.

We can now further condition our expression for $a^*_{ik|\ell j h}$ in \cref{eq:a-star-1} on $X(e)$, writing 
\begin{widetext}
\begin{align}
    a^*_{ik|\ell j h} 
    &= \sum_{s} \prob{S(e) = s \given I^*_{\ell j h}} \times\nonumber\\
    &\qquad\sum_{x} \prob{X(e) = x}  \prob{\abs{e\cap g} = k \given S(e) = s , X(e) = x, I^*_{\ell j h}} \prob{\abs{e} = i \given  S(e) = s ,  \abs{e\cap g} = k , X(e) = x, I^*_{\ell j h}} \\ 
    &= \sum_{s} \prob{S(e) = s \given I^*_{\ell j h}} \times\nonumber\\
    &\qquad\sum_{x} \gamma_x  \prob{\abs{e\cap g} = k \given S(e) = s , X(e) = x, I^*_{\ell j h}} \prob{\abs{e} = i \given  \abs{e\cap g} = k, S(e) = s ,  X(e) = x, I^*_{\ell j h}} \\ 
    &= \sum_{s} \underbrace{\prob{S(e) = s \given I^*_{\ell j h}}}_{t^{(1)}_{s|\ell jh}} \sum_{x} \gamma_x  \underbrace{\prob{\abs{e\cap g} = k \given S(e) = s , X(e) = x, I^*_{\ell j h}}}_{t^{(2)}_{k|sx\ell jh}} \underbrace{\prob{\abs{e} = i \given S(e) = s, X(e) = x}}_{t^{(3)}_{i|sx}}\;. \label{eq:a}
\end{align}
\end{widetext}
In the second line we have used the definition of $\vgamma$. In the third line we have used the fact that, conditional on $S(e)$ and $X(e)$, the size of $e$ is independent of the sizes of $f$, $g$, $f\cap g$, and $e\cap g$, because $S(e)$ and $X(e)$ specify the size of $e$ except for the novel nodes, which cannot intersect any other edges. 
We have also named the resulting terms, which we now proceed to compute. 

There are two relatively simple terms. First,
\begin{align}
    t^{(1)}_{s|\ell j h} = \prob{S(e) = s \given I^*_{\ell j h}} = \binom{\ell-1}{s-1}\eta^{s-1}(1-\eta)^{\ell-s}\;, \label{eq:t1}
\end{align}
since this is simply the probability of selecting a total of $s$ nodes from $f$ (which has size $\ell$) to form $e$ during the edge-sampling step.
Next, the term $t^{(3)}_{i|sx}$ is simply the probability of sampling $i - s - x$ from the novel node distribution: 
\begin{align}
    t^{(3)}_{i|sx} = \beta_{i-s-x}\;. \label{eq:t3}
\end{align}

The more complicated term is $t^{(2)}_{k|sx\ell jh}$. 
We condition on the value of $S(e,g)$:
\begin{widetext}
\begin{align}
    t^{(2)}_{k|sx\ell jh} &\eqdef \prob{\abs{e\cap g} = k \given S(e) = s, X(e) = x, I^*_{\ell j h}} \\ 
    &= \sum_{\sigma} \prob{\abs{e\cap g} = k \given S(e) = s, X(e) = x, S(e,g) = \sigma, I^*_{\ell j h}} \prob{S(e,g) = \sigma \given S(e) = s, X(e) = x, I^*_{\ell j h}} \\ 
    &= \sum_{\sigma} \underbrace{\prob{\abs{e\cap g} = k \given  X(e) = x, S(e,g) = \sigma, I^*_{\ell j h}}}_{t^{(4)}_{k|x\sigma \ell jh}} \underbrace{\prob{S(e,g) = \sigma \given S(e) = s,  I^*_{\ell j h}}}_{t^{(5)}_{\sigma | s \ell jh}} \;.
\end{align}
\end{widetext}
In the third line we have used two simplifications: first, $e\cap g$ depends on $e\cap f$ only through $e \cap f \cap g$, which is described by $S(e, g)$. 
Second, $S(e,g)$ is independent of $X(e)$, since $X(e)$ specifies the nodes in $e$ that are not in $e\cap f$.
We have also again named several terms which we will analyze further. 
First, $t^{(4)}_{k|x\sigma \ell jh}$ is the probability that $k - \sigma$ extant nodes are added to $e$ that are also added to $g$. 
There are $x$ total extant nodes added, $j - \sigma$ candidate extant nodes in $g$, and $n - \ell$ total candidate extant nodes. 
This probability is hypergeometric: 
\begin{align}
    t^{(4)}_{k|x\sigma \ell jh} &\eqdef \prob{\abs{e\cap g} = k \given  X(e) = x, S(e,g) = \sigma, I^*_{\ell j h}}  \\ 
    &= \mathrm{HyperGeometric}(k - \sigma; x, j - \sigma, n - \ell)  \\ 
    &= \frac{\binom{j - \sigma}{k - \sigma}\binom{n - \ell - j + \sigma}{x - k + \sigma}}{\binom{n-\ell}{x}}\;. \label{eq:t4}    
\end{align}
Unlike most of the other terms we have studied, this term includes a dependence on the extensive quantity $n$, which can be re-expressed (in expectation) in terms of $m$.
Let us parse the asymptotics of this term up to order $m^{-1}$. 
When $k < \sigma$, $t_{k|x\sigma\ell jh} = 0$. 
When $k = \sigma$, this expression simplifies in the asymptotic limit to 
\begin{align}
    t^{(4)}_{k|xk\ell jh} = \frac{\binom{n - \ell - j + k}{x}}{\binom{n-\ell}{x}} \doteq 1\;.
\end{align}
When $k = \sigma + 1$, we have 
\begin{widetext}
\begin{align}
    t^{(4)}_{k|x\sigma\ell jh} = \frac{\binom{j-\sigma}{1} \binom{n-\ell - j + \sigma}{x-1}}{\binom{n-\ell}{x}} \doteq (j-\sigma) \frac{{(n-\ell - j + \sigma)}^{x-1}}{(x-1)!} \frac{x!}{(n -\ell)^x} \doteq x(j-\sigma)n^{-1}\;. 
\end{align}
\end{widetext}
Recalling that $\mean{d}n = \mean{k}{m}$ and that $\frac{\mean{k}}{\mean{d}} = \mu_{\vbeta}$, we find that, when $k = \sigma+1$,  
\begin{align}
    t^{(4)}_{k|x\sigma\ell jh} = x(j-\sigma)\mu_{\vbeta} m^{-1}\;.
\end{align}
A similar calculation shows that, if $k > \sigma+1$, then $t^{(4)}_{k|x\sigma\ell jh} = O(m^{-2})$.
We therefore conclude 
\begin{align}
    t^{(4)}_{k|x\sigma\ell jh}  &\doteq \begin{cases}
        0 &\quad k < 0 \text{ or } k > j\\ 
        1 &\quad k = \sigma \\ 
        x(j-\sigma)\mu_{\vbeta} m^{-1} &\quad k = \sigma+1 \\ 
        O(m^{-2}) &\quad k > \sigma+1\;.
    \end{cases}
\end{align}

Next, $t^{(5)}_{\sigma|s\ell jh}$ is the probability that, among $s$ nodes added to $e$ through the edge-sampling step, a total of $\sigma$ of them are also elements of $g$.
This probability is also hypergeometric: we require $\sigma$ successful draws in $s$ total draws from a population of $\ell$ nodes in $f$ containing $\abs{f \cap g} = h$ ``successful'' nodes which are also elements of $g$. 
This gives 
\begin{align}
    t^{(5)}_{\sigma|s\ell jh} &\eqdef \prob{S(e,g) = \sigma \given S(e) = s, I^*_{\ell j h}}  \\ 
    &= \mathrm{HyperGeometric}(\sigma; s, h, \ell)  \\ 
    &= \frac{\binom{h}{\sigma}\binom{\ell - h}{s - \sigma}}{\binom{\ell}{s}}\;. \label{eq:t5}
\end{align}

These expressions then give us an approximation for $t^{(2)}_{k|sx\ell jh}$: we separate out the cases $\sigma = k$ and $\sigma = k-1$, yielding
\begin{widetext}
\begin{align}
    t^{(2)}_{k|sx\ell jh} &= \prob{\abs{e\cap g} = k \given S(e) = s, X(e) = x, I^*_{\ell j h}}  \\
    &= \sum_{\sigma} t^{(4)}_{k|x\sigma\ell jh} t^{(5)}_{\sigma|s\ell jh}  \\
    &\doteq t^{(4)}_{k|xk\ell jh} t^{(5)}_{k|s\ell jh} + t^{(4)}_{k|x(k-1)\ell jh} t^{(5)}_{k|s\ell jh} + O(m^{-2})  \\
    &= \frac{\binom{h}{k}\binom{\ell - h}{s - k}}{\binom{\ell}{s}} + \frac{\binom{h}{k-1}\binom{\ell - h}{s - k+1}}{\binom{\ell}{s}} x(j - k + 1)\mu_{\vbeta} m^{-1} + O(m^{-2})\\ 
    &\eqdef w^{(1)}_{k|s\ell h} + w^{(2)}_{k|sx\ell jh}m^{-1} + O(m^{-2})\;, \label{eq:t2}
\end{align}
\end{widetext}
where 
\begin{widetext}
\begin{align}
    w^{(1)}_{k|s\ell h} \eqdef \frac{\binom{h}{k}\binom{\ell - h}{s - k}}{\binom{\ell}{s}} \quad \text{and} \quad 
    w^{(2)}_{k|sx\ell jh} \eqdef \frac{\binom{h}{k-1}\binom{\ell - h}{s - k+1}}{\binom{\ell}{s}} x(j - k + 1)\mu_{\vbeta}\;. \label{eq:w1-and-w2}
\end{align}
\end{widetext}
This expression says that, to form an intersection of size $k$ with edge $g$, we either need to pick $k$ nodes from $g$ during the edge-sampling step, or $k-1$ nodes from $g$ during the edge-sampling step together with one additional node from the extant node addition step, with other possibilities being much less likely. 
Importantly, $w^{(1)}_{k|s\ell h} = 0$ iff $k \geq h+1$, while $w^{(2)}_{k|sx\ell jh} = 0$ iff $k \geq h+2$.
In particular, $w^{(2)}_{1|sx\ell j0} \geq 0$. 
Furthermore, $w^{(2)}_{k|sx\ell jh} = 0$ if $k = 0$. 

To sum up these calculations, we can insert our findings into \eqref{eq:a}:  
\begin{align}
    a^*_{ik|\ell jh } &\doteq \sum_{s} t^{(1)}_{s|\ell jh} \sum_{x} \gamma_x  t^{(2)}_{k|sx\ell jh} t^{(3)}_{i|sx} \\ 
    &\doteq\sum_{s} t^{(1)}_{s|\ell jh} \sum_{x} \gamma_x \paren{w^{(1)}_{k|s\ell h} + w^{(2)}_{k|sx\ell jh}m^{-1} + O(m^{-2})} t^{(3)}_{i|sx}  \\
    &\doteq \phi_{ik|\ell jh} + m^{-1}\psi_{ik|\ell jh} \;, 
\end{align}
where 
\begin{widetext}
\begin{align}
    \phi_{ik|\ell jh} &\eqdef \sum_{s} t^{(1)}_{s|\ell jh} \sum_{x} \gamma_x w^{(1)}_{k|s\ell h} t^{(3)}_{i|sx} \quad \text{and} \quad 
    \psi_{ik|\ell jh} \eqdef \sum_{s} t^{(1)}_{s|\ell jh} \sum_{x} \gamma_x w^{(2)}_{k|sx\ell jh} t^{(3)}_{i|sx}\;.
\end{align}
\end{widetext}
We note that, since  $w^{(2)}_{k|sx\ell jh} = 0$ if $k = 0$, it is also the case that $\psi_{ik|\ell jh} = 0$ if $k = 0$.
We also have that $\psi_{ik|\ell jh} = 0$ if $k \geq h + 2$ per the argument above. 
Our computations above imply that $\phi_{ik|\ell jh} = 0$ if $k > h$ or $k > \ell$, and $\psi_{ik|\ell jh} = 0$ if $k > h+1$,  $k > \ell$, or $k = 0$. 

Our approximate compartmental update now reads 
\begin{widetext}
\begin{align}
    p'_{ijk} &\doteq p_{ijk} + {\frac{1}{2}}b_{ik|j}\sum_{\ell h} \paren{r_{\ell j h} + r_{j\ell h}} + \frac{m-1}{2} \sum_{\ell h} \brackets{ \paren{\phi_{ik|\ell j h} + m^{-1} \psi_{ik|\ell j h}} r_{\ell j h} + \paren{\phi_{ik|j \ell h} + m^{-1} \psi_{ik|j \ell h}} r_{j \ell h}} \\ 
    &\doteq p_{ijk} + {\frac{1}{2}}b_{ik|j}\sum_{\ell h} \paren{r_{\ell j h} + r_{j\ell h}} + \frac{1}{2}\sum_{\ell h} \paren{\psi_{ik|\ell j h} r_{\ell jh} + \psi_{ik|j \ell h}r_{j \ell h}} + \frac{m-1}{2}\sum_{\ell h} \paren{\phi_{ik|\ell j h}r_{\ell j h} + \phi_{ik|j \ell h}r_{j\ell h}} \;. \label{eq:compartmental-linearized}
\end{align}
\end{widetext}

\subsubsection{Asymptotic Behavior of $\mathbf{r_{ijk}}$}

We now aim to use the compartmental update \cref{eq:compartmental-1}, along with the formulae in \cref{eq:zijk,eq:t1,eq:t3,eq:t2} to study the asymptotic behavior of $r_{ijk}$ as $m$ grows large.  

Let us assume that, at stationarity,  
\begin{align}
    r_{ijk} \doteq m^{-\lambda_k} q_{ijk} \label{eq:asymptotics}
\end{align}
for all $i, j, k$ where $q_{ijk}$ is a nonnegative constant independent of $m$, and $\lambda_k$ depends only on $k$ but not on $i$ or $j$. 
We assume that $q_{ijk} > 0$ when $k \leq i \land j$, provided that $i$ and $j$ are edge sizes supported by the model.  
Our aim is to determine the values of $\lambda_k$ and $q_{ijk}$. 
Our strategy is to substitute \cref{eq:asymptotics} into \cref{eq:compartmental-linearized} and then determine the values of $\lambda_k$ and $q_{ijk}$. 

Substituting \cref{eq:asymptotics} into \cref{eq:compartmental-linearized}, along with $p'_{ijk} = \binom{m+1}{2}r_{ijk}$ and $p_{ijk} = \binom{m}{2}r_{ijk}$ gives 
\begin{widetext}
\begin{align}
    \binom{m+1}{2}(m+1)^{-\lambda_k}q_{ijk} &\doteq \binom{m}{2}m^{-\lambda_k}q_{ijk} + \frac{1}{2}b_{ik|j} \sum_{h\leq \ell} m^{-\lambda_h} (q_{\ell jh} + q_{j\ell h}) + \frac{1}{2}\sum_{\ell, h } m^{-\lambda_h}\paren{\psi_{ik|\ell j h} q_{\ell jh} + \psi_{ik|j \ell h}q_{j \ell h}} \nonumber\\
        &\quad  + \frac{m-1}{2}\sum_{\ell, h} m^{-\lambda_h} \paren{q_{\ell j h} \phi_{ik|\ell jh} + q_{j \ell h} \phi_{ik|j\ell h}}\;.
\end{align}
We now move $\binom{m}{2}m^{-\lambda_k}q_{ijk}$ to the left side and simplify 
\begin{align}
    \binom{m+1}{2}(m+1)^{-\lambda_k}q_{ijk} - \binom{m}{2}m^{-\lambda_k}q_{ijk} &= q_{ijk}\brackets{\frac{m(m+1)}{2}(m+1)^{-\lambda_k} - \frac{m(m-1)}{2}m^{-\lambda_k}} \\ 
    &= \frac{q_{ijk}}{2}\brackets{m(m+1)^{1-\lambda_k} - (m-1)m^{1-\lambda_k}} \\ 
    &= \frac{m^{1-\lambda_k} q_{ijk}}{2} \brackets{m\paren{\frac{m+1}{m}}^{1-\lambda_k} - (m-1)} \\ 
    &= \frac{m^{2-\lambda_k} q_{ijk}}{2} \brackets{\paren{\frac{m+1}{m}}^{1-\lambda_k} - \frac{m-1}{m}} \\ 
    &= \frac{m^{2-\lambda_k} q_{ijk}}{2} \brackets{1 + \frac{1}{m}(1-\lambda_k) +  o\paren{\frac{1}{m}} - 1 + \frac{1}{m}} \\ 
    &= \frac{m^{2-\lambda_k} q_{ijk}}{2} \brackets{(2-\lambda_k)\frac{1}{m} + o\paren{\frac{1}{m}}} \\ 
    &\doteq \frac{m^{2-\lambda_k} q_{ijk}}{2} \brackets{(2-\lambda_k)\frac{1}{m}} \\ 
    &= \frac{(2-\lambda_k)m^{1-\lambda_k} q_{ijk}}{2} \\ 
    &\triangleq c_k m^{1-\lambda_k} q_{ijk}\;,
\end{align}
where we have defined $c_k = \frac{2-\lambda_k}{2}$.
We assume throughout that $\lambda_k \neq 2$.
We then have 
    \begin{align}
        c_k m^{1-\lambda_k} q_{ijk} &\doteq \frac{1}{2}b_{ik|j} \sum_{h\leq \ell} m^{-\lambda_h} (q_{\ell jh} + q_{j\ell h}) + \frac{1}{2}\sum_{\ell, h} m^{-\lambda_h}\paren{\psi_{ik|\ell j h} q_{\ell jh} + \psi_{ik|j \ell h}q_{j \ell h}} \nonumber\\ 
        &\quad + \frac{m-1}{2}\sum_{\ell, h} m^{-\lambda_h} \paren{q_{\ell j h} \phi_{ik|\ell jh} + q_{j \ell h} \phi_{ik|j\ell h}} \\ 
        &\doteq \frac{1}{2}b_{ik|j} \sum_{h\leq \ell} m^{-\lambda_h} (q_{\ell jh} + q_{j\ell h}) + \frac{1}{2}\sum_{\ell, h} m^{-\lambda_h}\paren{\psi_{ik|\ell j h} q_{\ell jh} + \psi_{ik|j \ell h}q_{j \ell h}} \nonumber\\ 
        &\quad + \frac{1}{2}\sum_{\ell, h} m^{1-\lambda_h} \paren{q_{\ell j h} \phi_{ik|\ell jh} + q_{j \ell h} \phi_{ik|j\ell h}}\;,
    \end{align}
yielding 
    \begin{align}
        q_{ijk} &\doteq \frac{1}{2c_k}b_{ik|j} \sum_{h \leq \ell} m^{\lambda_k-\lambda_h-1} (q_{\ell jh} + q_{j\ell h}) + \frac{1}{2c_k}\sum_{\ell, h} m^{\lambda_k-\lambda_h-1}\paren{\psi_{ik|\ell j h} q_{\ell jh} + \psi_{ik|j \ell h}q_{j \ell h}} \nonumber\\ 
        &\quad + \frac{1}{2c_k}\sum_{\ell, h} m^{\lambda_k-\lambda_h} \paren{q_{\ell j h} \phi_{ik|\ell jh} + q_{j \ell h} \phi_{ik|j\ell h}} \label{eq:asymptotics-3}
    \end{align}
\end{widetext}
provided that $\lambda_k \neq 2$. 

We now determine the values of $\lambda_k$. 
We first consider $k = 0$. 
In this case, $b_{ik|j} = 0$ and $\psi_{ik|\ell jh} = 0$ for all $i$ and $j$. 
This simplifies \cref{eq:asymptotics-3} to 
\begin{align}
    q_{ij0} \doteq \frac{1}{2c_0}\sum_{\ell, h} m^{\lambda_0-\lambda_h} \paren{q_{\ell j h} \phi_{i0|\ell jh} + q_{j \ell h} \phi_{i0|j\ell h}}\;.
\end{align}
The requirement that $q_{ij0}$ be a constant implies that either (a) $a_{i0|\ell jh}=0$ and $a_{i0|\ell jh}=0$ or (b) $\lambda_0 \leq \lambda_h$ for all $h \geq 0$. 
Case (a) occurs only when no novel nodes are added to the hypergraph ($\beta_0 = 1$). 
For the remainder of this section, we will assume that this is not the case. 
In case (b), the normalization requirement 
\begin{align}
    1 &= \sum_{ijk} r_{ijk} = \sum_{ijk} m^{-\lambda_k}q_{ijk}\;.
\end{align}
implies that $\lambda_k = 0$ for at least one choice of $k$.
It follows that $\lambda_0 = 0$. 

Furthermore, since $b_{ik|j} > 0$ whenever $k \leq i\land j$, \cref{eq:asymptotics-3} implies that $\lambda_k - \lambda_h - 1 \leq 0$. 
Choosing $h = 0$ implies that $\lambda_k \leq 1$ for all $k$.
We now show that, under our assumptions, $\lambda_k = 1$ for all $k \geq 1$. 
Fix $k = \argmin_{h \geq 1}\lambda_h$. 
Consider the exponent of $m$ in the first two terms of \cref{eq:asymptotics-3}, which is $\lambda_k - \lambda_h - 1$ as $h$ ranges.  
Let us first assume that these terms do not vanish as $m\rightarrow \infty$. 
This requires that $\lambda_k = 1 + \lambda_{h^*}$ for at least one choice  $h = h^*$.
If $h^* \geq 1$, we may repeat this argument to find $h^{**}$ such that $\lambda_{h^*} = 1 + \lambda_{h^{**}}$. 
But then, $\lambda_k = 2 + \lambda_{h^{**}}$, which contradicts the requirement that $\lambda_k - \lambda_{h^{**}} - 1 \leq 0$. 
Therefore, we must have $h^* = 0$, from which it follows that $\lambda_k = 1$. 

So far, we have shown that, for any $k$ such that the first two terms of \cref{eq:asymptotics-3} do not vanish, we must have $\lambda_k = 1$.
We will now show that if the two terms of \cref{eq:asymptotics-3} do vanish for some $k = k^*$, then $q_{ijh} = 0$ for all $i$,  $j$, and $h \geq k^*$.
Since the sum defining the first two terms includes $h = 0$, vanishing of the first two terms would imply that $\lambda_{k^*} < 1$. 
In order for the second term to remain bounded, we must also have $\lambda_h \geq \lambda_{k^*}$ for all $h \geq k^*$, with at least one $h\geq k^*$ such that $\lambda_h = \lambda_{k^*}$.
Indeed, the second term of \cref{eq:asymptotics-3} can be maximized by setting $\lambda_{h} = \lambda_{k^*}$ for all $h \geq {k^*}$.
This gives the approximate linear system 
\begin{align}
    q_{ijk} \doteq \frac{1}{2c_k} \sum_{\ell, h}  \paren{q_{\ell j h} \phi_{ik|\ell jh} + q_{j \ell h} \phi_{ik|j\ell h}} \quad \forall k \geq k^*\;. \label{eq:linear-system-contradiction}
\end{align}
Recalling that $\psi_{ik|\ell jh} = 0$ whenever $k > h$, we find that this system is closed in the entries $q_{ijk}$ such that $h \geq {k^*}$, and implies that the vector $\mathbf{q}$ is an eigenvector with eigenvalue 1 of the matrix $\mathbf{C}$ whose action on $\mathbf{q}$ is defined by \cref{eq:linear-system-contradiction}. 
This, however, is impossible, since $\phi_{ik|\ell jh} \geq 0$ for all $i, k, \ell, j, h$, $\sum_{\ell, h }\phi_{ik|\ell jh} = 1$, and $c_k \geq \frac{1}{2}$, which means that $\frac{1}{c_k}\sum_{\ell, h }\phi_{ik|\ell jh} < 1$ and therefore $\frac{1}{2c_k}\sum_{\ell, h }\brackets{\phi_{ik|\ell jh} + \phi_{ik|j \ell h}} < 1$.
This implies that the spectral radius of $\mathbf{C}$ is strictly less than 1, so the only solution to the system is $\mathbf{q} = \mathbf{0}$.
This contradicts the assumption that $q_{ijk} > 0$ for all $i$, $j$, and $k\geq k^*$.
We conclude that the first term of \cref{eq:asymptotics-3} does not vanish for any $k$.

Summarizing, \emph{under our assumptions}, 
\begin{align}
    \lambda_k = \begin{cases} 
        0 &\quad k = 0 \\ 
        1 &\quad k \geq 1\;.
    \end{cases}
\end{align}
To complete our asymptotic analysis, it is necessary to describe the values of $q_{ijk}$. 
We proceed from \cref{eq:asymptotics-3}. 
When $k = 0$, $\lambda_k = 0$ and $c_k = 1$. We then have 
\begin{widetext}
\begin{align}
    q_{ij0} &\doteq \frac{1}{2}\sum_{\ell, h\geq 0} m^{\lambda_0-\lambda_h} \paren{q_{\ell j h} \phi_{i0|\ell jh} + q_{j \ell h} \phi_{i0|j\ell h}} \\ 
    &= \frac{1}{2}\brackets{\sum_{\ell} \paren{q_{\ell j 0} \phi_{i0|\ell j0} + q_{j \ell 0} \phi_{i0|j\ell 0}} + \sum_{\ell, h\geq 1} m^{-1} \paren{q_{\ell j h} \phi_{i0|\ell jh} + q_{j \ell h} \phi_{i0|j\ell h}}}  \\
    &\doteq \frac{1}{2}\sum_{\ell} \paren{q_{\ell j 0} \phi_{i0|\ell j0} + q_{j \ell 0} \phi_{i0|j\ell 0}}\;. \label{eq:system-1}
\end{align}
Next, when $k \geq 1$, $\lambda_k = 1$ and $c_k = \frac{1}{2}$. We then have 
\begin{align}
    q_{ijk} &\doteq \frac{1}{2c_k}b_{ik|j} \sum_{h\leq \ell} m^{\lambda_k-\lambda_h-1} (q_{\ell jh} + q_{j\ell h}) + \frac{1}{2c_k}\sum_{\ell, h \leq k} m^{\lambda_k-\lambda_h-1}\paren{\psi_{ik|\ell j h} q_{\ell jh} + \psi_{ik|j \ell h}q_{j \ell h}} \nonumber\\
    &\quad + \frac{1}{2c_k}\sum_{\ell, h\geq k} m^{\lambda_k-\lambda_h} \paren{q_{\ell j h} \phi_{ik|\ell jh} + q_{j \ell h} \phi_{ik|j\ell h}} \\ 
    &\doteq b_{ik|j} \sum_{h\leq \ell} m^{\lambda_k-\lambda_h-1} (q_{\ell jh} + q_{j\ell h}) + \sum_{\ell, h \leq k} m^{\lambda_k-\lambda_h-1}\paren{\psi_{ik|\ell j h} q_{\ell jh} + \psi_{ik|j \ell h}q_{j \ell h}} \nonumber\\
    &\quad + \sum_{\ell, h\geq k} m^{\lambda_k-\lambda_h} \paren{q_{\ell j h} \phi_{ik|\ell jh} + q_{j \ell h} \phi_{ik|j\ell h}} \\ 
    &= b_{ik|j} \sum_{h\leq \ell} m^{-\lambda_h} (q_{\ell jh} + q_{j\ell h}) + \sum_{\ell, h \leq k} m^{-\lambda_h}\paren{\psi_{ik|\ell j h} q_{\ell jh} + \psi_{ik|j \ell h}q_{j \ell h}} \nonumber\\
    &\quad + \sum_{\ell, h\geq k}  \paren{q_{\ell j h} \phi_{ik|\ell jh} + q_{j \ell h} \phi_{ik|j\ell h}} \\ 
    &\doteq b_{ik|j} \sum_{\ell} (q_{\ell j0} + q_{j\ell 0}) + \sum_{\ell} \paren{\psi_{ik|\ell j 0} q_{\ell j0} + \psi_{ik|j \ell 0}q_{j \ell 0}} + \sum_{\ell, h\geq k} \paren{q_{\ell j h} \phi_{ik|\ell jh} + q_{j \ell h} \phi_{ik|j\ell h}}\;.
\end{align}
In the case $k=1$, this becomes 
    \begin{align}
        q_{ij1} &\doteq b_{i1|j} \sum_{\ell}  (q_{\ell j0} + q_{j\ell 0}) +\sum_{\ell} \paren{\psi_{i1|\ell j 0} q_{\ell j0} + \psi_{i1|j \ell 0}q_{j \ell 0}} +  \sum_{\ell, h\geq 1} \paren{q_{\ell j h} \phi_{i1|\ell jh} + q_{j \ell h} \phi_{i1|j\ell h}} \label{eq:system-k1}\;,
    \end{align}
while the $k>1$ case simplifies further, using the fact that
$\psi_{ik|\ell jh} = 0$ for $k \geq h+2$, 
to give
    \begin{align}
        q_{ijk} &\doteq b_{ik|j} \sum_{\ell} (q_{\ell j0} + q_{j\ell 0}) + \sum_{\ell, h\geq k}  \paren{q_{\ell j h} \phi_{ik|\ell jh} + q_{j \ell h} \phi_{ik|j\ell h}} \label{eq:system-2} \quad\mathrm{for}\quad k>1\;.
    \end{align}
\end{widetext}
Jointly, \cref{eq:system-1,eq:system-k1,eq:system-2} define an approximate linear system for $q_{ijk}$: 
\begin{align}
    \mathbf{q} = \mathbf{C} \mathbf{q}\;,
\end{align}
where the entries of $\mathbf{C}$ are defined to appropriately conform to the entries of a (vectorized) $\mathbf{q}$. 
We note that $\mathbf{C}$ has nonnegative entries, so $\mathbf{q}$ has to be its Perron eigenvector. 
In principle, writing down $\mathbf{C}$ and finding the Perron eigenvector would be sufficient to determine $\mathbf{q}$. 

\subsection{Computational Challenges}

In the experiment shown in the main text, we tracked edge sizes and intersections up to size $12$, resulting in a matrix $\mathbf{C}$ of size $12^3 \times 12^3$.
Experimentally, we found that the LAPACK solver (accessed through the \texttt{numpy.linalg.eig} function in Python) was able to accurately find the leading eigenpair for this matrix for some but not all parameter combinations, with larger values of $\eta$ especially leading to convergence issues. 
Our experimental evidence suggests that this was indeed a solver issue rather than an issue with our proposed approximation scheme, in that the numerically obtained solution in such cases was demonstrably not an eigenvalue. 
Because we considered the results such as those shown in the main text to constitute sufficient evidence of the correctness of our approximation scheme, we did not pursue other solvers or otherwise attempt to solve the system for all parameter combinations.

\section{Descriptions of Alternative Models} \label{sec:alternative-models}

\subsection{Growing Hypergraph Erd\H{o}s-R\'enyi model}

We list the steps that we take for growing hypergraphs in an Erd\H{o}s-R\'enyi manner in a framework similar to what we did for HCM in the main text. We emphasize that our approach here is only one way of generating hypergraphs that generalize Erd\H{o}s-R\'enyi networks and Preferential Attachment networks; this particular approach was chosen to maintain a roughly consistent edge size with the \model at each time step. Recall that at each time step of the \model, we seed the new edge $e$ with nodes from existing edge $f$ (first uniformly sampling one node from $f$, and then adding each other node IID with probability $\eta$) --- we denote this positive integer as $\alpha$. The \model step also includes $g \sim \extantnodedist$ extant nodes and $b \sim \novelnodedist$ novel nodes. Denoting the newly formed edge as $\at{e}{t+1}$, our corresponding Erd\H{o}s-R\'enyi generalization proceeds as follows.

\begin{enumerate}
    \item \textbf{Extant node sampling}: Following the above notation, we select $\alpha + g$ nodes drawn uniformly at random without replacement from $\at{\nodeset}{t}$ to initiate $\at{e}{t+1}$.
    \item \textbf{Novel node addition}: We next sample $\hat{b}$ from a Poisson distribution with mean $b$ (so that there is some randomness in this number compared to the \model) and add $\hat{b}$ novel nodes to $\at{e}{t+1}$.
\end{enumerate}
After forming $\at{e}{t+1}$, we have an Erd\H{o}s-R\'enyi update $\at{\hypergraph}{t+1} = (\at{\nodeset}{t} \cup \braces{\at{e}{t+1}}, \at{\edgeset}{t} \cup {\at{e}{t+1}})$. 

\subsection{Hypergraph Preferential Attachment Model}

We do not employ the model of \cite{avinRandomPreferentialAttachment2019} due to lack of available code for simulation or inference. Instead, the generalization of preferential attachment we use here takes the different numbers from the \model step, using the two different numbers of extant nodes to mix preferential and uniform selection. Note that the \textbf{Novel node addition} step below is exactly the same as described for Erd\H{o}s-R\'enyi above.
\begin{enumerate}
    \item \textbf{Degree based sampling}: We select $\alpha$ nodes drawn with probability proportional to their degree without replacement from $\at{\nodeset}{t}$ and name this set $\alpha'$.
    \item \textbf{Extant node sampling}: We select $g$ nodes drawn uniformly at random without replacement from $\at{\nodeset}{t} \setminus \alpha'$ to initiate $\at{e}{t+1}$.
    \item \textbf{Novel node addition}: We next sample $\hat{b}$ from a Poisson distribution with mean $b$ (so that there is some randomness in this number compared to the \model) and add $\hat{b}$ novel nodes to $\at{e}{t+1}$. 
\end{enumerate}
After forming $\at{e}{t+1}$, we have a Preferential Attachment update $\at{\hypergraph}{t+1} = (\at{\nodeset}{t} \cup \braces{\at{e}{t+1}}, \at{\edgeset}{t} \cup {\at{e}{t+1}})$.

\end{document}